\documentclass[aps,rmp,twocolumn,superscriptaddress,noshowkeys,amsmath,amssymb,amsfonts]{revtex4-1}
\usepackage[utf8x]{inputenc}
\usepackage{bm}
\usepackage{dcolumn}
\usepackage{graphicx}
\usepackage{color}
\usepackage{tikz}
\usepackage{soul}

\newcommand{\tikzcircle}[2][black,fill=black]{\tikz[baseline=-0.5ex]\draw[#1,radius=#2] (0,0) circle ;}
\newcommand{\tikzcirclew}[2][black,fill=white]{\tikz[baseline=-0.5ex]\draw[#1,radius=#2] (0,0) circle ;}%

\begin{document}
\title{{\textit Colloquium}: Ice Rule and  
Emergent Frustration in Particle Ice and Beyond}
\author{Antonio Ortiz-Ambriz}
\affiliation{Departament de F\'isica de la Mat\`eria Condensada Universitat de Barcelona, Barcelona, Spain.}
\affiliation{Institut de Nanoci\`encia i Nanotecnologia (IN$^2$UB) Universitat de Barcelona,  Barcelona, Spain.}
\author{Cristiano Nisoli}
\affiliation{Theoretical Division and Center for Nonlinear Studies
Los Alamos National Laboratory Los Alamos, New Mexico 87545 USA}
\author{Charles Reichhardt}
\affiliation{Theoretical Division and Center for Nonlinear Studies
Los Alamos National Laboratory Los Alamos, New Mexico 87545 USA}
\author{Cynthia J. O. Reichhardt}
\affiliation{Theoretical Division and Center for Nonlinear Studies
Los Alamos National Laboratory Los Alamos, New Mexico 87545 USA}
\author{Pietro Tierno}
\affiliation{Departament de F\'isica de la Mat\`eria Condensada Universitat de Barcelona, Barcelona, Spain.}
\affiliation{Institut de Nanoci\`encia i Nanotecnologia (IN$^2$UB) Universitat de Barcelona,  Barcelona, Spain.}
\affiliation{Universitat de Barcelona Institute of Complex Systems (UBICS) Universitat de Barcelona, Barcelona, Spain.}
\email{ptierno@ub.edu}
\date{\today}
\begin{abstract}
Geometric frustration and the ice rule are two concepts 
that are intimately connected and widespread across condensed matter.
The first refers to the inability of a system to satisfy
competing interactions in the presence of spatial constraints.
The second, in its more general sense, represents
a prescription for the minimization of the topological charges 
in a constrained system. Both can lead to  manifolds of high susceptibility and non-trivial, constrained disorder where exotic behaviors can appear and even be designed deliberately.
In this Colloquium, we describe the emergence of geometric frustration and the ice rule in soft condensed matter. This Review excludes the extensive developments of mathematical physics within the field of geometric frustration, but rather focuses on systems of confined micro- or mesoscopic particles that emerge as a novel paradigm exhibiting spin degrees of freedom.
In such systems, geometric frustration
can be engineered
artificially by controlling the spatial topology and geometry of the lattice,
the position of the individual particle units, or their relative filling fraction.
These capabilities enable the creation of
novel and exotic
phases of matter, and also
potentially lead
towards technological
applications
related to memory and logic devices that are based on 
the motion of topological defects. We review the rapid progress in theory and experiments and discuss the intimate physical connections with other frustrated systems at different length scales.
\end{abstract}
\maketitle
\tableofcontents{}
\section{Introduction}
Frustration in life emerges with the impossibility of simultaneously satisfying a set of requirements. Frustration in physics is not very different. A classical example is that of three spins on the vertex of a triangle that want to be antiferromagnetically aligned~\cite{wannier1950}. This requirement cannot be realized all around the triangle, so at least two spins will display ferromagnetic order and will generate one frustrated bond. More generally, a geometrically frustrated system is one that is subjected to local requirements that cannot be satisfied collectively along certain loops in the system. The notion is therefore  intrinsically topological, i.e. invariant under transformations that do not  rip those frustrated loops, and indeed it lends itself to abstract, elegant treatments in term of Wilson loops and gauge symmetry, at least in the case of spin systems~\cite{fradkin1978gauge}. 
In real systems, however, the local requirement is usually the minimization of an energy as a pairwise interaction that is in general geometry-dependent. 
  
Frustration is essential for the understanding of a variety of real materials, such as spin glasses~\cite{mydosh2014spin}, water ice~\cite{bernal1933theory,Pauling1935}, spin ices~\cite{Ramirez1999,Bramwell2001,harris1997geometrical}, and even systems at different length scales such as granular materials ~\cite{richard2005slow}, liquid crystals~\cite{kamien2001,Lopez-Leon2011}, filament bundles~\cite{Grason2016a}, coupled lasers~\cite{Nixon2013} and many others~\cite{Grason2016b}. There, it is often---but by no means always---associated with slow relaxation, degeneracy, and zero temperature entropy. Indeed, frustration implies compromise, and therefore produces various forms of so-called {\it constrained disorder}, which are manifolds whose disorder obeys some non trivial rules, either local or global.  The ice rule is an important
example of a local rule that constrains disorder
and that can have global, {\it i.e.} topological, implications. 

Typically in an ensemble characterized by constrained disorder, violations of  local rules appear as localized excitations of the low-energy states of the system. These local rule violations control the collective dynamics. An example, described further below, is provided by the magnetic monopoles in spin ices~\cite{ryzhkin2005magnetic,Castelnovo2008a}, which are violations of an ice manifold. Another example are the emergent  topological charges in certain artificial spin ices~\cite{lao2018classical}. These concepts are
of both practical interest, such as for the measurement of currents of magnetic
monopoles~\cite{Giblin2011}, and also of theoretical interest.
Indeed, we are
accustomed to understanding
topological defects in terms of
alterations of an underlying order, such as misplaced books on a library shelf or dislocations in a crystal, rather than in terms of disorder. 

Manifolds of constrained disorder can be rich playgrounds for new physics, and frustration is a fundamental ingredient in the design of artificial systems that can generate novel exotic behaviors which are often not found in natural materials~\cite{Wang2006, Nisoli2013,heyderman2013artificial}.
Artificial spin ice systems were first created
by mimicking natural frustrated geometries
and by realizing celebrated models of statistical mechanics~\cite{lieb1967residual,Baxter1982}
in settings that allowed characterization at the constituent level,
often in real time.
However, since artificial materials can be realized in various geometries,
a more recent effort~\cite{Morrison2013,nisoli2017deliberate} has
advanced
the design of new systems generating a wide variety of new phenomena,
including
dimensionality reduction, emergent classical topological order, realizations of Pott's models, phase transitions, ice rule fragility, and quasi-crystal spin ices~\cite{gilbert2016emergent,gilbert2014emergent,lao2018classical,sklenar2019field,gliga2017emergent,barrows2019emergent,shi2018frustration,louis2018tunable,ostman2018interaction,ma2016,Libal2018,perrin2016extensive}. Furthermore, many of these ideas proved to be exportable across different platforms, from nanomagnets to trapped colloids, to liquid crystals, and to superconductors~\cite{Ortiz-Ambriz2016,libal2009,Latimer2013,Wang2018,Duzgun2019Artificial}. 

\section{The ice rule}
We start by briefly recalling the history of the ice rule, and its appearance in magnetic systems either natural or lithographically fabricated. For a more extensive treatment we refer the reader to available reviews and commentaries on the subject:
\cite{Ramirez1994,Bramwell2001,Gardner2010,Nisoli2013,heyderman2013artificial,gilbert2016frustration,nisoli2018frustration}. 

\subsection*{1. From water ice to spin ice}
\begin{figure*}
\includegraphics[width=0.9\textwidth]{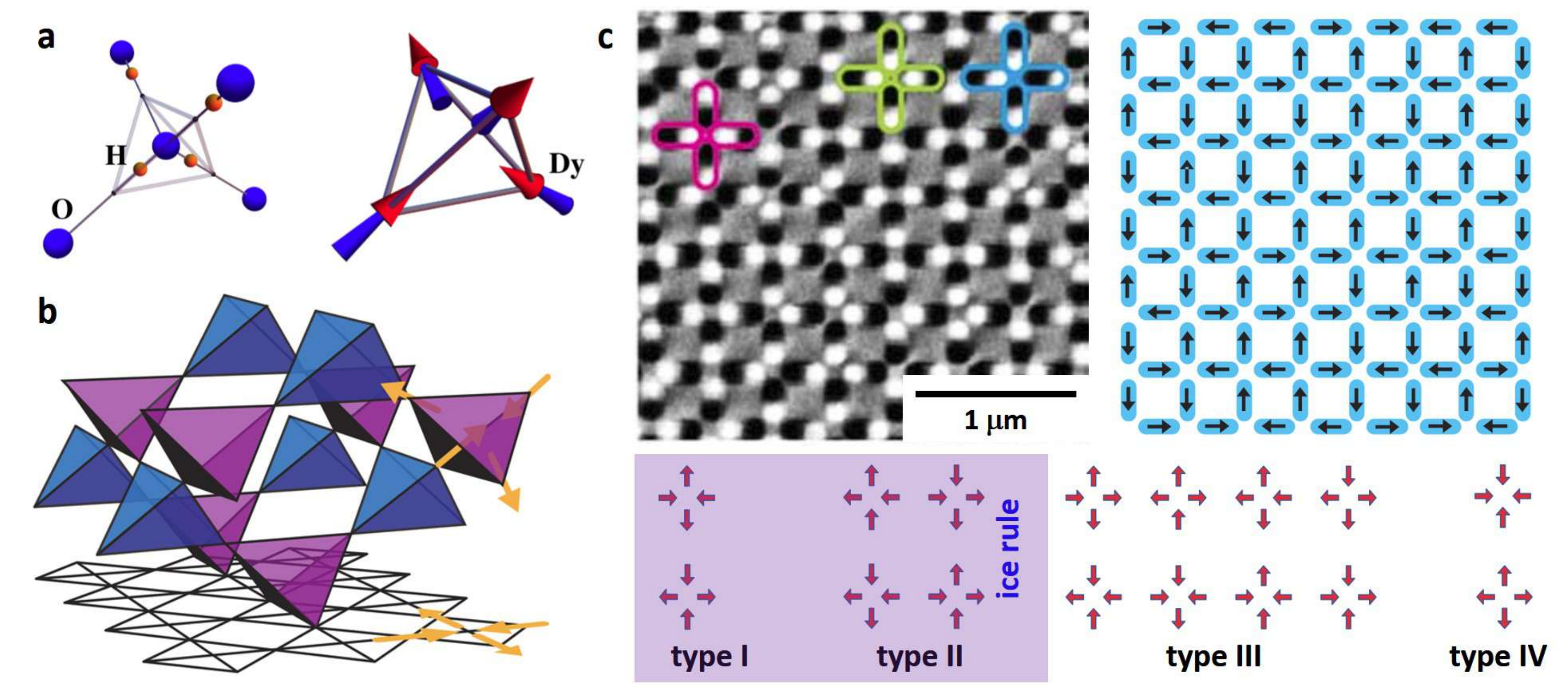}
\caption{(a) Ice rule for the water ice I$_h$ (left) and magnetic spin ice material (right). The first is characterized by oxygen atoms sharing protons and located at the vertices of a diamond lattice. The spin ice has Ising-like moments pointing along the same lattice,
from \textcite{Castelnovo2008a}.   
(b) A projection of the three-dimensional pyrochlore lattice onto a two-dimensional plane gives rise to a square lattice, from \textcite{Glaetzle2014}.
(c) Left: Magnetic force microscope image of an artificial spin ice composed by a square lattice of permalloy islands with lattice spacing of $400\rm{nm}$. Right: corresponding spin orientations. Bottom: vertex configurations with ice rule highlighted in the violet box. 
Image edited with permission from \cite{Wang2006}.}
\label{Fig_1}
\end{figure*}

Many fundamental properties of water are still not completely understood~\cite{chaplin2006we}.
One of
water's early mysteries pertained to its residual entropy and was solved by Linus Pauling in the 1930's.
Through a series of carefully conducted calorimetric experiments,
Giaque and Ashley~\cite{giauque1933molecular,Gia36}
had found that the entropy of ice at low temperature was not zero. Pauling~\cite{Pauling1935} explained it via the ice rule of Bernal and Fowler~\cite{bernal1933theory}. In water ice each oxygen atom is bound to four others via hydrogen atoms.
Two of these hydrogen atoms
are close to the oxygen atom in covalent bonds, and
two are further away forming covalent bonds with neighboring oxygens,
Fig.~\ref{Fig_1}(a), left.
The number of ways in which
the hydrogen atoms can be arranged, Pauling showed,
grows exponentially with the number of oxygen
atoms, leading to a non-zero entropy per molecule.  
Ice-like systems received renewed interest in the 1990's
with the discovery a class of magnetic substances which were named ``spin ices'' because their magnetic texture at low temperature presents a degenerate low energy state with a residual entropy,  consistent with the ice rule model. These rare earth titanates, such as Ho$_2$Ti$_2$O$_7$ and Dy$_2$Ti$_2$O$_7$, presented frustrated interactions between the magnetic moments of their constituent at low temperature. Their magnetic cations Ho$^{3+}$ and Dy$^{3+}$ were  known to carry a very large magnetic moment, about ten times the Bohr magneton.
At low temperatures
such moments could be regarded as
binary, classical Ising spins constrained to point along the directions of the lattice bonds
forming the  pyrochlore lattice,
as shown in Fig.~\ref{Fig_1}(a), right.
It was then noted~\cite{harris1997geometrical}  and confirmed experimentally~\cite{Ramirez1999} that the  resulting ferromagnetic interaction favors the ice rule, in which two spins point in each vertex and two out, in analogy with the allocation of protons in the lattice of water ice. 

\subsection*{2. Artificial spin ice systems}

Between the late 1990's and the early 2000's,  a mature effort in the exploration of the magnetic state of nanodots and nanoislands~\cite{bader2006colloquium} focused on  how to obtain exotic magnetic textures and transitions. 
Meanwhile,  a different
approach to exotic behavior
involved {\it relatively simple} elongated, single-domain, magnetic nanoislands,
whose magnetization could be described by an Ising pseudo-spin.
In this case, complexity is introduced via the
{\it mutual interaction} of these nanoislands in order
to generate possibly interesting collective states of exotic emergent behaviors~\cite{Wang2006,Tanaka2006, Nisoli2013,heyderman2013artificial,Libal2006a}. This approach provided a twofold advantage. First, characterization
of the individual degrees of freedom in real space
is possible via methods such as Magnetic Force Microscopy (MFM), Photoelectron emission microscopy (PEEM), Transmission Electron Microscopy (TEM), Surface Magneto-Optic Kerr Effect (MOKE), and Lorentz Microscopy.
Secondly, the collective behavior of these systems is open to design.

The resulting so-called Artificial Spin Ices (ASI) could mimic the behavior of spin ice rare earth pyrochlores---hence the name---but at desirable temperature and field ranges. In a certain geometry, ASI  represents a plane projection of a pyrochlore material, as shown in Figs.~\ref{Fig_1}(b,c). Soon, a  growing number of groups began using ASI to investigate topological defects, the dynamics of magnetic charges, and spin fragmentation~\cite{Mengotti2010,Ladak2010,lad2011,zeissler2013non,phatak2011nanoscale,ladak2011direct,pollard2012propagation,Rougemaille2011,canals2016},
as well as information encoding~\cite{Lammert2010,Wang2016b}, equilibrium and nonequilibrium thermodynamics~\cite{Nisoli2007, Nisoli2010,Ke2008,cugliandolo2017artificial,levis2013thermal,Morgan2011,budrikis2012disorder,budrikis2011diversity,Lammert2012,Nisoli2012,chioar2014kinetic,chioar2014kinetic}, avalanches~\cite{hugli2012emergent,mellado2010dynamics}, direct realizations of the Ising system~\cite{zhang2012perpendicular,arnalds2016new,nisoli2016nano, chioar2014nonuniversality, chioar2016ground}, magnetoresistance and the Hall effect~\cite{Branford2012,PhysRevB.95.060405}, critical slowing down~\cite{Anghinolfi2015}, dislocations~\cite{drisko2017topological}, spin wave excitations~\cite{gliga2013spectral}, ratchet effects~\cite{gliga2017emergent}, dimensionality reduction~\cite{gilbert2016emergent}, classical topological states~\cite{gilbert2014emergent,lao2018classical,perrin2016extensive}, quasi-crystals~\cite{barrows2019emergent,shi2018frustration},  and memory effects~\cite{Gilbert2015a,Libal2012a}.

\subsection*{3. Ice rule and topology: conceptual themes}

The ice-rule appears an innocuous enough concept, yet it can be understood in most general terms and has profound implications. 
Consider a lattice or even a directed graph,
with binary variables such as Ising spins on each edge,
impinging in vertices of various coordination $z$. The pyrochlore or square geometries introduced before can serve as examples. Then we can define the  {\it topological charge} of a vertex of coordination $z$ with $n$ spins pointing toward as
\begin{equation}
q=2n-z,
\label{charge}
\end{equation}
which corresponds to the difference between
the number of spins pointing in and out. This notion is properly topological as it only depends on the topology of the graph. Furthermore, given a spin configuration, any single spin flip will alter the charge in two nearby vertices. Only flipping  proper loops of spins will preserve the charge distribution.  

In this language, the ice rule can be considered a prescription for a local minimization of $|q|$  at each vertex. In practical systems this is typically (but not necessarily) enforced by the nearest neighboring spin-spin interactions. Any configuration of spins that locally minimizes $|q|$ is said to obey the ice rule, and the subset of the phase space corresponding to such configurations is called an {\it ice manifold}.
For a lattice of
uniform, even coordination, the ice manifold is then characterized by zero charge  $q=0$ on each vertex. The first violation of the ice rule then corresponds to charges $q=\pm2$, called  monopoles. Very often--but not always--
 the ice manifold represents the lowest energy of the system, and then the ground state is degenerate and disordered, and has a residual entropy. 
Importantly it cannot be ``explored from within." Any single spin flip on an ice-rule configuration creates a
pair of monopoles of opposite charge $\pm 2$~\cite{ryzhkin2005magnetic,Castelnovo2008a}, violating the ice rule.  Only the coherent flipping of a loop of spins, properly chosen so that they are all arranged head to tail, represents an ``update" that does not violate the ice rule. 

The simplest, square version of such system has motivated theoretical research in applied mathematics for half a century. In the 1960s, Lieb, Wu, Baxter, Rys, and others began working on simplified, two-dimensional models of mathematical physics, known as vertex models,
that captured and also generalized the properties of ice. In these models, different energies are assigned to different vertex configurations on a square lattice, and in many cases the models can be solved exactly, typically via transfer matrix methods~\cite{lieb1967residual,Wu1969,Baxter1982, rys1963ueber,Lie67}. 
The six-vertex model in particular~\cite{lieb1967residual} only admits ice rule obeying vertices on a square lattice, and was meant to represent a solvable, two-dimensional equivalent of water ice. As it forbids monopoles, it is completely embedded into the  ice manifold  and has strong topological properties. For instance, if the degeneracy of the ice-manifold is lifted by an energetics that selects the antiferromagnetic state, as in the Rys F-model~\cite{rys1963ueber},  then the corresponding ordering transition is infinitely continuous~\cite{Lie67},  {\it and}  with an order parameter, also infinitely continuous!~\cite{Baxter1982}. 

Beside clarifying the topological nature of the ice rule, these models  were also often equivalent to other important
statistical mechanics systems, such as dimer cover models,
and thus initiated an independent theoretical line of research in mathematical physics that has further evolved in terms of loop representations to describe topological effects at (and of) the boundaries.
We will not report here on this half a century long, very interesting developments because it goes beyond the scope of this Review. 
Indeed the interesting phenomenology in those models arises from a topological structure---given by the ice rule---that in simulations and experiments is always violated. While it is true that  such topological structure is present in the ground state, this does not imply that it extrapolates fully to the low energy physics, which is in fact generally a dynamics of topological defects. 
In a sense, realistic spin ice systems are neither topologically constrained nor unconstrained. They should perhaps be called "Topology Breaking" systems because for any non-zero temperature, violations of the topological constraints (the ice-rule) in form of monopoles substantially changes the physics. For a dramatic example: while the antiferromagnetic ground state of artificial square ice~\cite{Wang2006,Nisoli2010,Porro2013,Zhang2013,Morgan2011} would seem to be well described by the Rys F-model, the ordering transition  of the latter is in the Kosterlitz Thouless class~\cite{kosterlitz1973ordering,Lie67}, whereas in ``real'' square ice the transition is simply second order in the Ising class~\cite{Wu1969,levis2013thermal,sendetskyi2019continuous} precisely because monopoles break the topological constraint.

Thus in real spin ice systems it becomes less interesting to eviscerate all the possible topological representations of their idealized ice manifold. It is instead more interesting to see how the topological properties of an ultimately unreachable  ground state  affect the low-energy physics as the system  breaks those topological constraints. 
This issue has been attacked via the concepts of spin {\it fractionalization}  into magnetic charges~\cite{ryzhkin2005magnetic,Castelnovo2008a} and of spin {\it fragmentation}~\cite{petit2016observation} into a Coulomb~\cite{henley2010coulomb} and non-Coulomb magnetization.
For instance, in pyrochlore ice the ice manifold can be considered as a classical topological phase of constrained disorder~\cite{henley2010coulomb, henley2011classical,castelnovo2012a}.
Such a phase is labeled not by an order parameter, as an ordered phase would, but rather by a fluctuating, solenoidal gauge field $\vec M ( x)$, which can be thought of as a coarse grain of the spin magnetization. Then spin fractionalization implies that the low-energy manifold can be described in terms of local violations of its solenoidal nature, such that
\begin{equation}
\vec \nabla \cdot \vec M(x) =4\pi \sum_{i} q_{x_i}\delta(x-x_i).
\end{equation}
Thus, the ice manifold corresponds to $q=0$ everywhere, and it is easy to show~\cite{henley2010coulomb} that it implies algebraic
(specifically, dipolar) correlations in the reciprocal space, and thus the observed pinch points in the structure factor of neutron scattering. While this ice phase is  critical,  this is a purely mathematical abstraction since no transition to the ice manifold exists, and at non-zero temperature monopoles will always be present, however sparsely, and provide a correlation length. 

The low energy of the system remains however reminiscent of the topological nature of its ground state. Through spin fragmentation~\cite{petit2016observation} it is possible to show that the low-energy ensemble can be decomposed into an ice-rule ensemble plus an crystal-charge ensemble. In terms of the coarse grained magnetization this merely corresponds to an Helmholtz decomposition of the magnetization vector in the sum of a divergence free and divergence full part. Because the two components are uncoupled at the lowest order in the effective free energy, dipolar correlations and pinch points survive at non-zero temperature. 
Artificial realization of topological phases was achieved in the Shakti geometry~\cite{lao2018classical}
as well as in magnetic square ice~\cite{perrin2016extensive,Moller2006} and rectangular~\cite{ribeiro2017realization,loreto2018experimental} ice.
No equivalent phase has yet been realized with colloids, though those magnetic realization provide directions. 

For vertices of odd coordination, the cancellation of the charge in Eq.~(\ref{charge})
 is impossible and $|q|$ is minimized when $q=\pm1$. Thus in a simple system of uniform,
odd coordination, such as the kagom{\' e} ice~\cite{Qi2008}, the ice manifold corresponds to ensembles of disordered spins  obeying a 2-in/1-out or 2-out/1-in ice-rule at each vertex.
Because there is no charge cancellation, such systems can be regarded
as disordered, neutral plasma of opposite topological charges which, depending on realization, can interact, leading to further phases within the ice manifold~\cite{Moller2009,Chern2011,Chern,Zhang2013,drisko2015fepd,Libal2018a}, as we will show below in more detail. 

Finally, systems of mixed coordination, say for instance $z=3,4$ host both versions of the ice rule, 2-in/2-out on $z=4$ coordinated vertices and 2-in/1-out or 2-out/1 on $z=3$ vertices. Here is where the difference between magnetic and particle-based spin ice becomes more dramatic, with the ice-rule breaking down in the latter, as we shall see in Section IV.  

\section{The ice rule in soft (particle) systems}

\subsection*{1. Colloids as a model system}
Colloidal particles represent a versatile model system to explore collective phenomena in statistical physics and condensed matter, due to their accessible length-scales and to the presence of simple and tunable interactions \cite{Ana03,Poo04}. 
Colloidal systems have provided new insight into
the glass transition \cite{Weeks2017}, 
yielding phenomena \cite{Schall2007}, and the motion of dislocations \cite{Schall2004a}, and
have been used to test numerous foundations of both equilibrium \cite{Zahn1999,Thorneywork2017b} 
and nonequilibrium statistical mechanics \cite{Martinez2014,Martinez2017}.
One of the best examples of the
use of colloids for both
condensed matter and statistical physics
studies is in understanding ordering and commensuration effects
on surfaces,
where the substrate can be one-dimensional (1D) \cite{Bechinger2001},
two-dimensional (2D) \cite{Brunner2002,Mangold2003,Bohlein2012b},
quasiperiodic \cite{Mikhael2008}, or
random \cite{Deutschlander2013}. 
For colloids interacting with 
periodic 1D surfaces,
various effects such as transitions among
liquid, 2D hexagonal solid, and smectic states appear
as a function of increasing substrate strength \cite{Bechinger2001,Tierno2008,Tierno2012}. 
The next level of complexity is
to consider colloids interacting with two dimensional periodic arrays, such as
an egg carton \cite{Mangold2003,Bohlein2012b,Reichhardt2002,Sarlah2005,Agra2004}, muffin tin \cite{Bechinger2001}, or more complex potentials~\cite{Yellen2005,Gunnarsson2005,Grzybowski2013,Tierno2014,Tierno20142,Daniel2016,Loehr2016,Massana2019}.
Such systems can mimic the ordering of atoms on 2D surfaces \cite{Coppersmith1982}, 
vortices in type-II superconductors with nanostructured pinning \cite{Baert1995,Martin1999,Harada1996}, 
and vortices in Bose-Einstein condensates \cite{Tung2006} interacting with 2D
optical trap arrays.
In this case, commensuration effects arise when the
number of colloids is an integer multiple of the number of
potential minima, giving an integer filling factor $f$, where at $f=1$ each
trap captures a single colloidal particle. 
Experiments \cite{Bechinger2001,Bohlein2012b} and theoretical studies \cite{Reichhardt2002,Sarlah2005,Agra2004}
of colloids interacting with 2D substrates
reveal a variety of novel 
orderings, including commensurate colloidal molecular crystals for $f = 2,3...N.$ 
When the system is away from commensuration,
such as just above $f = 1.0$,
the additional particles
act like highly mobile kinks,
and both simulations and experiments for colloids
on 2D arrays have revealed
the motion of these kinks 
under an applied drive \cite{Bohlein2012b,Vanossi2012,McDermott2013a}.
Other studies of colloids have focused on the Aubry transitions that occur
as a function of substrate strength \cite{Brazda2018}.           

Since various substrates for colloidal
particles can be created readily,
including systems in which a
single trap contains
a double well potential \cite{Babic2005}, 
a natural question to ask is whether it is possible to create a colloidal
system containing
effective spin degrees of freedom.
For colloids on regular 2D arrays, it was shown that for
integer $f = 2$ and
higher, the colloids in each trap
act like dimers with a spin degree of freedom that
can be mapped to an Ising one,
making it possible to use
colloids in periodic substrate arrays
to model various spin systems \cite{Reichhardt2002,Sarlah2005,Agra2004}.

\begin{figure}
\includegraphics[width=\columnwidth]{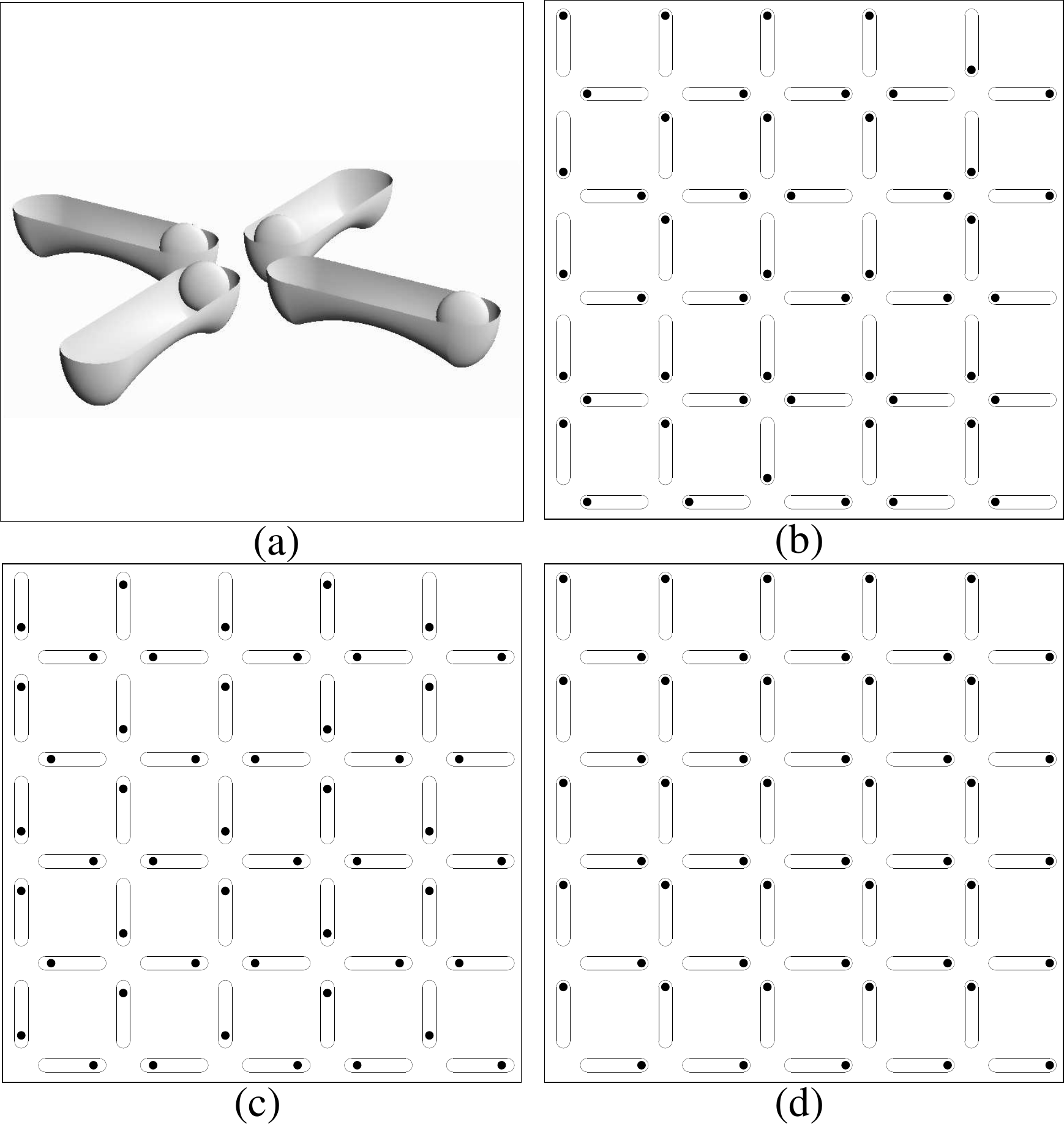}
\caption{
(a)  Schematic  of the  basic  unit cell  with four double wells each filled by one
colloid. (b-d) Images of a small portion  of  the system for different electrostatic charges $q_e$. Dark  circles: colloids;  ellipses: traps.   
(b) A random   vertex   distribution at $q_e=0$.
(c) A long range  ordered  square  ice  ground  state at $q_e=1.3$.
(d) A biased system at $q_e=0.4$ where a drive of $F^{dc}=0.02$ has been applied
at an angle of $45$ degrees from the $\hat{\bm{x}}$ axis, from \textcite{Libal2006a}.
}
\label{fig:1}
\end{figure}

\subsection*{2. Artificial colloidal ice: simulations}

Motivated by these ideas,
and influenced by the work on
magnetic ASI,
Lib{\' a}l {\it et al.}
proposed that 2D array of double well traps for colloids could mimic 
a square ASI \cite{Libal2006a}.
In this case the basic unit cell consists of
four double well traps arranged as  shown in Fig.~\ref{fig:1}(a), where
$f=1$ so that each trap captures a single colloidal particle.
The colloid-colloid interaction potential is of
Yukawa or screened Coulomb form, $V(r) \propto q_e\exp(-\kappa r)/r$,
where $q_e$ is the electrostatic charge and $1/\kappa$ is the screening length.

The simulations for $N$ charged colloids on a 2D array of $N$ traps
are performed using
2D Brownian dynamics (BD).
The colloid dynamics are overdamped
and the equation of motion for colloid $i$ is:
\begin{equation}
\eta\frac{ d{\bf r}_{i}}{dt} = {\bf F}_{i}^{cc} + {\bf F}^{T}_{i} + 
{\bf F}^{ext}_{i} + {\bf F}^{s}_{i} 
\end{equation}
where in rescaled units the damping constant is set to $\eta=1.0$ and
$a_0$ is used as the unit of distance in the simulation.
The colloid-colloid interaction force is given by,
${\bf F}_{i}^{cc} = -F_0\sum^{N}_{i\neq j}\nabla_i V(r_{ij})$
with
$V(r_{ij}) = (1/r_{ij})\exp(-\kappa r_{ij}){\bf {\hat r}}_{ij}$.
Here $r_{ij}=|{\bf r}_{i} - {\bf r}_{j}|$,
${\bf {\hat r}}_{ij}=({\bf r}_{i}-{\bf r}_{j})/r_{ij}$,
${\bf r}_{i(j)}$ is the position of particle $i$($j$),
$F_0=q_e^{2}/(4\pi\epsilon\epsilon_0)$,
$\epsilon$ is the
solvent dielectric constant, and $\kappa=4/a_0$  which is the typical screening length for charged colloidal systems.
These simulations
neglect hydrodynamic interactions between the colloids, which is a reasonable
assumption
for charged particles in the low volume fraction limit or where the dynamics are 
dominated by hopping events rather than large scale flows.
The effects of thermal noise are captured in the simulation via 
the force term ${\bf F}^T$,
which represents random Langevin kicks with the
properties $\langle{\bf F}^{T}_{i}\rangle = 0$ and
$\langle {\bf F}_i^{T}(t) {\bf F}_j^{T}(t^{\prime})\rangle = 
2\eta k_{B}T\delta_{ij}\delta(t - t^{\prime})$.
Unless otherwise mentioned, $F^T=|{\bf F}^{T}|=0$.

For charged colloidal systems, a bias can be applied using an electric
field in order to mimic
the effect of an external magnetic field in 
magnetic spin ice.
In the simulation model, this biasing field is represented by the term
${\bf F}^{\rm ext}_{i}$.
If, for example, the field
${\bf F}^{\rm ext}_{i} = F^{dc}(\hat{\bm{x}} + \hat{\bm{y}})$ is used to bias the
colloids along a 45 degree angle from the $\hat{\bm{x}}$ axis, the ice state
illustrated in Fig.~\ref{fig:1}(d) emerges.
\begin{table}
\begin{tabular}{|c|c|c||c|c|c|}
\hline
Type & Configuration & $E_i/E_{III}$ & Type & Configuration & $E_i/E_{III}$\\
\hline
I & 0000 & 0.001 & IV & 1001 & 7.02\\
II & 0001 & 0.0214 & V & 1101 & 14.977\\
III & 0101 & 1.0 & VI & 1111 & 29.913\\
\hline
\end{tabular}
\caption{Normalized electrostatic energy $E_i/E_{III}$ 
for
each vertex type. 
An example configuration for each vertex is
listed; 1 (0) indicates a colloid close to (far from) the vertex, 
from \textcite{Libal2006a}.
}
\end{table}

In the absence of a substrate, charged colloids
confined to two dimensions will form a triangular lattice~\cite{Kusner1994}.
For the traps in Fig.~\ref{fig:1}(a), there are two
equally favorable locations.
The two resulting configurations
can be mapped to a spin degree of freedom
in which the spin is defined to point toward the end of the trap occupied
by the colloid.
The lowest energy state for the single unit cell shown in Fig.~\ref{fig:1}(a)
has all the colloids sitting at the end of the trap that is furthest from the
vertex, so that all of the effective spins point away from the vertex.  This
minimizes the colloid-colloid interaction energy.
The highest energy state has all of the colloids sitting close to the vertex,
so that all of the effective spins point toward the vertex.
This corresponds
to a doubly charged monopole state.

As shown in
Fig.~\ref{fig:1}(b), when many unit cells are assembled into a lattice, it is not possible for all vertices to adopt their lowest energy
configuration simultaneously since 
the geometric arrangement competes with the symmetry of the pair interaction.
If the electrostatic screening is very strong or the trapping sites are
sufficiently far enough apart, the colloids do not interact with each other and
the arrangement of the effective spins is random,
as illustrated in
Fig.~\ref{fig:1}(b)
where the populations of the different possible vertex states
matches with what is expected for a purely thermal distribution. 
For strong colloid-colloid interactions,
the system forms 
the lowest energy collective ground state, in which
each vertex has two colloids close to it and two colloids far from it, equivalent to the
ice-rule of the square ASI (2-in/2-out), Fig.~\ref{fig:1}(c).
If a biasing field is applied at $45^{\circ}$ along the $\hat{\bm{x}}$ axis,
the spin ice rule is still obeyed
but the system forms
a biased state with high energy vertices which still obey the ice rule, Fig.~\ref{fig:1}(d).

The system can be characterized according to the six possible vertex types,
whose energy is listed in Table I.
From the lowest to the highest energy, the vertex types are:
type I, with all four spins out to create a double monopole;
type II, with one spin in and three spins out to create a monopole;
type III, the spin ice rule obeying state;
type IV, the biased ice rule state;
type V, with three spins in and one spin out to create a monopole;
and type VI, with all four spins in to create a high energy double monopole.
An immediate difference from the magnetic version of square ASI
is that in the colloidal artificial ice, the monopole states
II and V, and the double monopole states I and VI have different energies.
This 
becomes important in the mobility of
the monopoles, and is an indication of the fact
that the particle based spin ice system
minimizes
the global interaction energy rather than the local vertex energy. 

\begin{figure}[h]
\includegraphics[width=\columnwidth]{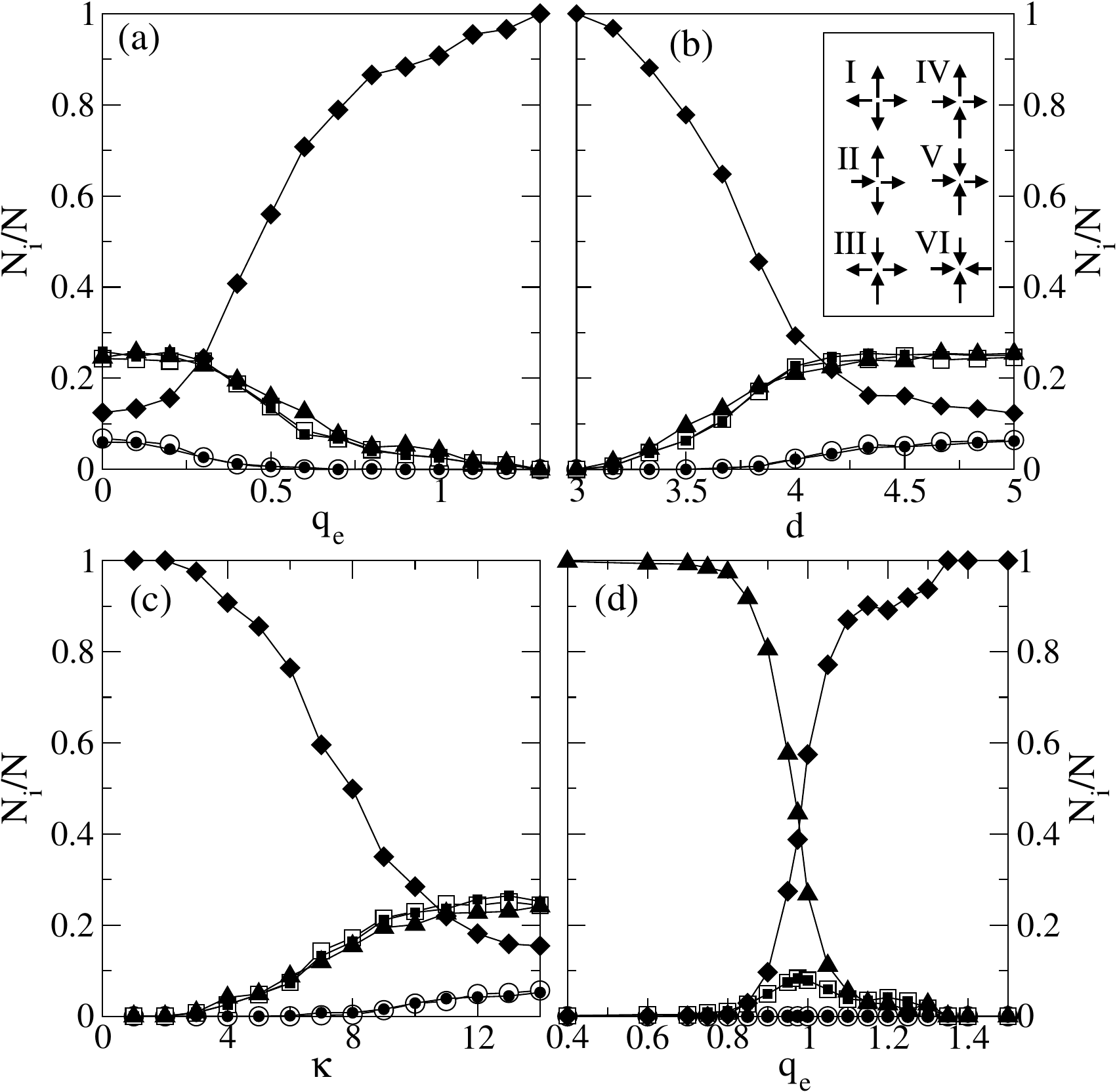}
\caption{  
$\bigcirc$: $N_I/N$; $\square$ : $N_{II}/N$; $\blacklozenge$: $N_{III}/N$;
$\blacktriangle$: $N_{IV}/N$; $\blacksquare$: $N_{V}/N$; $\bullet$:
$N_{VI}/N$.
(a) $N_i/N$ vs $q_e$ at $d=3$ and $\kappa=4.0$.
(b) 
$N_i/N$ vs $d$ at $q_e=1.3$ and $\kappa=4.0$.
Inset: schematic spin representation of the 6 vertex types.
(c) $N_i/N$ vs $\kappa$ at $d=3$ and $q_e=1.0.$
(d) $N_i/N$ vs $q_e$ for a biased system at $d=3$, $\kappa=4.0$,
and $F^{dc}=0.02$, from \textcite{Libal2006a}.
}
\label{fig:2}
\end{figure}

Using numerical simulations of this geometry,
\textcite{Libal2006a} examined 
the evolution of the vertex populations when passing
from the weak to the strong interacting limit.
Figure~\ref{fig:2}(a) shows 
the fraction $N_{i}/N$ of the $N$ vertices
that are of type $i$ as a function of
colloidal charge $q_e$
in a system with 
fixed $\kappa$ and distance between the traps.
At $q_e = 0$, 
the fractions of the vertex types match what is expected in a
non interacting system, while as $q_e$ increases,
type I and VI vertices disappear first followed by type II and IV vertices,
until for $q_e\geq 2.0$,
only the type III ice rule obeying vertices remain.
Figure~\ref{fig:2}(b) illustrates
the fraction of vertex types as a function of the distance $d$ between the traps in
a system with fixed $q_e$ and $\kappa$.
When $d$ is small, the colloids are strongly coupled and type III vertices dominate,
while as $d$ increases, the vertex distribution gradually shifts back
to the random configuration.
The results in Fig.~\ref{fig:2}(b) are very similar to
the observations
in the initial work on magnetic ASI,
where increasing the spacing between magnetic islands 
produced a more random state \cite{Wang2006}.
In a system with fixed $q_e$ and $d$ and changing screening strength $\kappa$,
shown in Fig.~\ref{fig:2}(c),
the interaction between the colloids is weak for strong screening and the
vertex distribution becomes random.
A sample with fixed
$d$ and $\kappa$ at
$q_e = 0.4$ that is subjected to an additional biasing drive $F^{dc}$ at 45 degrees
from the $x$ axis is initially
in the type IV biased state.
As $q_e$ increases, Fig.~\ref{fig:2}(d) shows that
a transition occurs into
the lower energy type III square ice state.
\textcite{Libal2006a} 
also found that a transition
from an  ordered ice state to a disordered state
occurs as a function of increasing temperature.
These results indicate
that a particle based artificial ice system
exhibit the same ice rule
obeying states as nanomagnetic ASIs,
with the advantages of the mesoscopic character of the particles
which
makes the microscopic degrees of freedom readily accessible.
%
\begin{figure}[t]
\begin{center}
\includegraphics[width=\columnwidth,keepaspectratio]{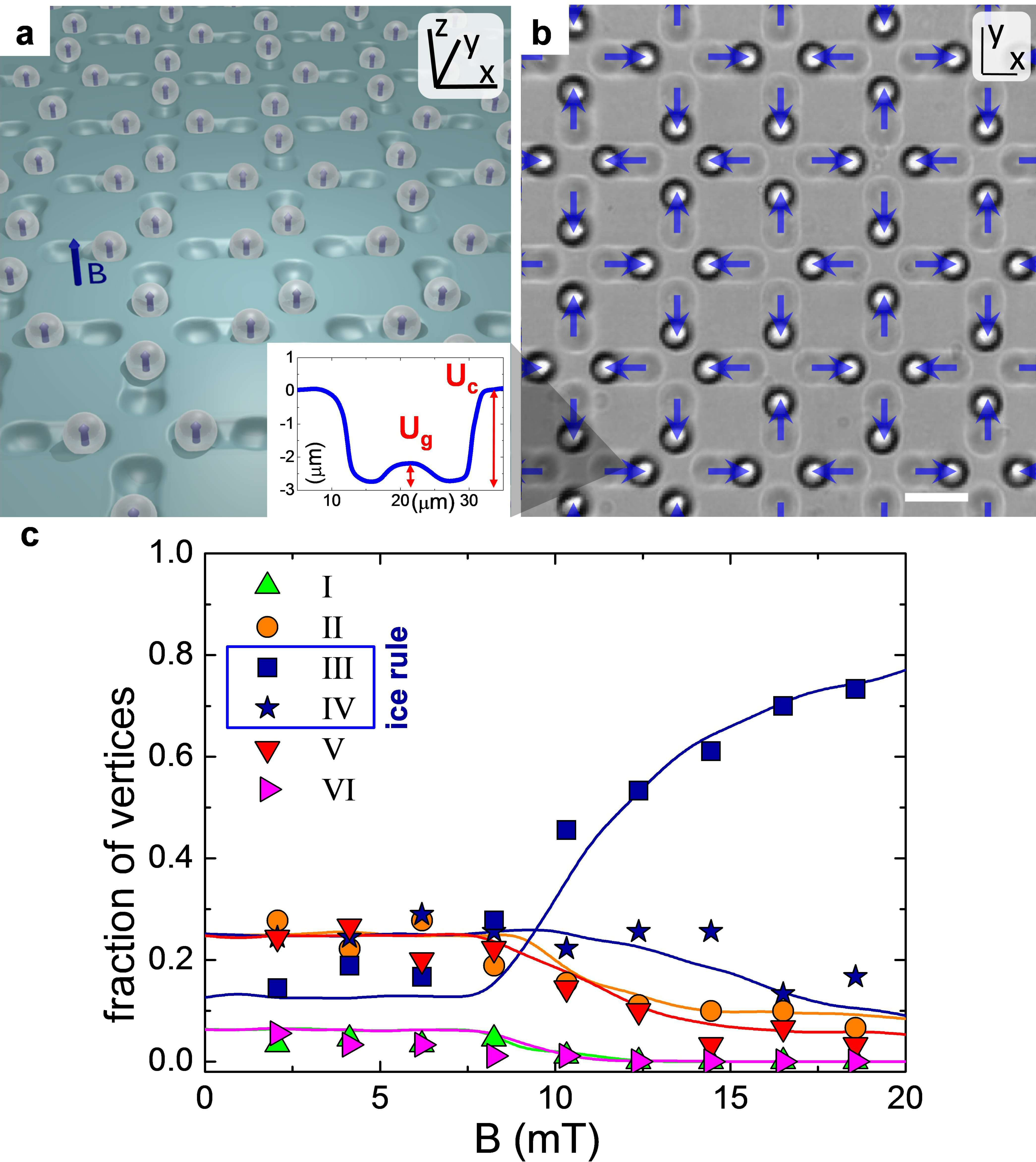}
\caption{(a) Schematic of a colloidal ice on a square lattice. Blue arrows in the particles are induced moments due to the applied field $\bm{B}$. Inset at the bottom shows a cross-section of one double well obtained via confocal microscopy, from~\textcite{Loehr2016b}. (b) Microscope image of the experimental system in the ($\hat{x},\hat{y}$) plane. Here, blue arrows indicate the spin associated with the particles, and the scale bar is $20 \mu m$. (c) Fraction of vertices for square ice versus applied field. Experimental data are scattered points, continuous lines are numerical simulations,
adapted from~\textcite{Ortiz-Ambriz2016}.}
\label{figure1}
\end{center}
\end{figure}

\subsection*{3. Experimental realization}

The realization of a full optical system,
following the original proposition  \cite{Libal2006a}, remains a challenging task because (a) it  would require a large optical power to generate enough double wells and (b) it is  difficult to finely tune DLVO based electrostatic interactions between colloidal particles. 
An alternative realization was demonstrated recently  
using a combination of different techniques including soft-lithography,
magnetic manipulation, and optical tweezers \cite{Ortiz-Ambriz2016}. 
The experimental system, illustrated in Figs.~\ref{figure1}(a,b), features interacting paramagnetic colloids confined by gravity in lithographically generated
topographic double wells.
Each trap contains two deep wells connected by a small central hill,
and these indentations are arranged in a regular lattice,
such as the square ice geometry illustrated in Fig.~\ref{figure1}(a). 
Using optical tweezers, each trap is filled with one paramagnetic colloid
consisting of a spherical polymer particle that is responsive to magnetic fields. 
Repulsive and tunable interactions are induced by applying an external magnetic field $\bm{B}=B\hat{\bm{z}}$, perpendicular to the particle plane. 
The applied field induces a dipole moment $\bm{m} =\pi d^3  \chi \bm{B}/(6\mu_0)$ within the particles, where $d$ is the particle diameter,
$\chi$ is the magnetic volume susceptibility, and
$\mu_0$ is the permeability of the medium (water). All particles interact through
magnetic dipolar interactions,
Pairs of particles $(i,j)$ with moments $\bm{m}_{i,j}=m \bm{e}_{i,j}$ and at distance $r=|\bm{r}_i-\bm{r}_j|$ interact through dipolar forces, with an interaction potential given by, 
$U_d(r) = \omega 
\left[\frac{\bm{e}_i \cdot \bm{e}_j}{r^3} - \frac{3(\bm{e}_i \cdot \bm{r})(\bm{e}_j \cdot \bm{r})}{r^5} \right]$, with $\omega = \mu_0 m^2/(4 \pi)$.
This potential is maximally attractive (repulsive) for particles with magnetic moments parallel (perpendicular) to $\bm{r}$.
In an unconstrained system, when the applied field is perpendicular to the particle plane, $U_d$ reduces to an isotropic repulsion between parallel particles in the same plane, $U_d(r) = \omega/r^3$. The field amplitude is chosen such that a confined particle, when subjected to the dipolar force from a neighbor,
can cross the central hill with gravitational potential $U_g$, but never escape from the bistable confinement ($U_c$), $U_g <U_d < U_c$, 
as shown in the small inset in Fig.~\ref{figure1}(a). 

Following the original idea~\cite{Libal2006a}, one can assign a vector to each particle pointing towards the well occupied by the colloid,
Fig.~\ref{figure1}(b).
This mapping makes it possible to construct a set of vertex rules similar to ASIs.
However, in contrast to ASI whose islands have in-plane dipole moments, the colloidal ice features out-of-plane dipoles, and thus the energetic hierarchy is similar to
that found for repulsive electrostatic colloids.
Within a homogeneous lattice, collective interactions between the particles oppose
the local energetics and enforce the ice rule for the colloidal system \cite{Nisoli2014}.
Indeed, systematic measurements combined with Brownian dynamics simulations confirm that, for high field amplitudes, the colloidal ice follows the ice rule  \cite{Ortiz-Ambriz2016}.
The experiments were performed by setting the system in a random configuration
using optical tweezers, turning on the magnetic interactions, and waiting for the system to reach thermal equilibrium.
The initial disordering process was necessary since, in contrast to the original simulations \cite{Libal2006a}, the magnetic colloids experience
negligible thermal fluctuations due to the  
relatively large particle size, $d\sim 10 \mu m$. The disordering process
can be considered as a way to couple the system with a virtual heat bath,
while increasing the strength of the pair interaction represents
a cooling procedure. 
Moreover, in analogy with ASIs, 
the particle-based ice
reaches a unique ground state (GS) 
filled by type III vertices. The loss of 
degeneracy results from
the different
particle distance at the vertex,
which makes the type III and type IV vertices energetically different. 
The most energetically unfavorable vertices
with three (type V) or four (type VI) colloids {\it in}
become topologically connected to the low energy
type II and type I vertices,
reducing their occurrence at large $\bm{B}$,
as shown in Fig.~\ref{figure1}(c).

\subsection*{4. Defect dynamics, grain boundaries, and logic gate}

The experimental system provides a versatile approach for probing the effect of disorder and topological defects in frustrated lattices. The optical tweezers, which are independent from the collective magnetic coupling, can be used to manually add or remove particles from the double wells.
%
\begin{figure}[t]
\begin{center}
\includegraphics[width=\columnwidth,keepaspectratio]{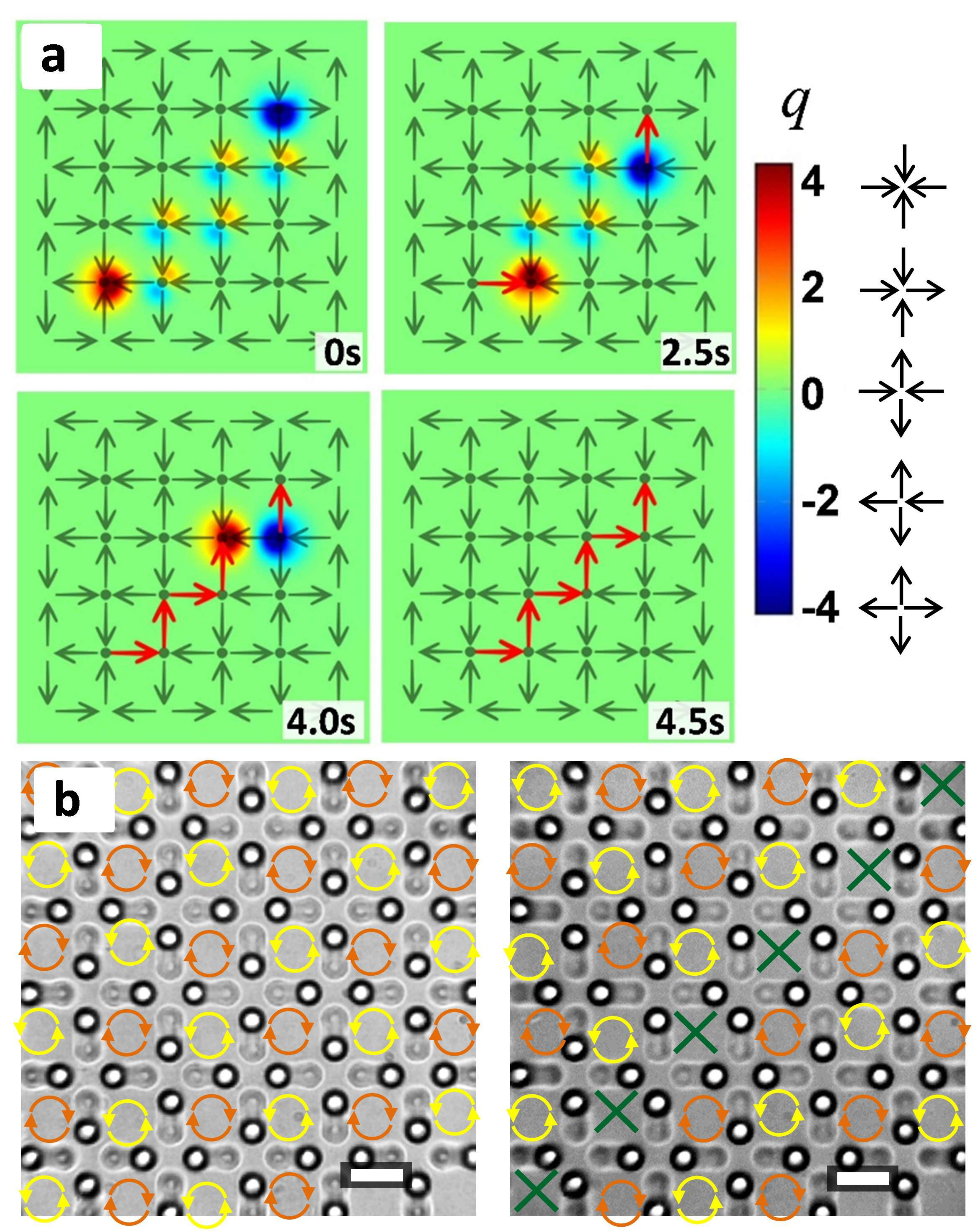}
\caption{(a) Colormap showing the vertex charges for a line of flipped particles connecting two $q = \pm 2$ 
defects. These defects annihilate after $t=4.5$ s due to the external field,
applied at $t=0$ s.
Image edited with permission from \textcite{Loehr2016b}.
(b) Microscope images showing a square ice state prepared in its ground state of type IV ($q=0$) vertices (left), or in a metastable state with a grain boundary (right). The square plaquettes can have 
a clockwise (orange arrows) or counterclockwise (yellow arrows) chirality (type IV vertices), or be achiral (green crosses). Scale bars are $20 \mu$m for both cases. Image edited with permission from \textcite{Ortiz-Ambriz2016}.}
\label{figure2}
\end{center}
\end{figure}
%
\begin{figure*}[t]
\begin{center}
\includegraphics[width=\textwidth,keepaspectratio]{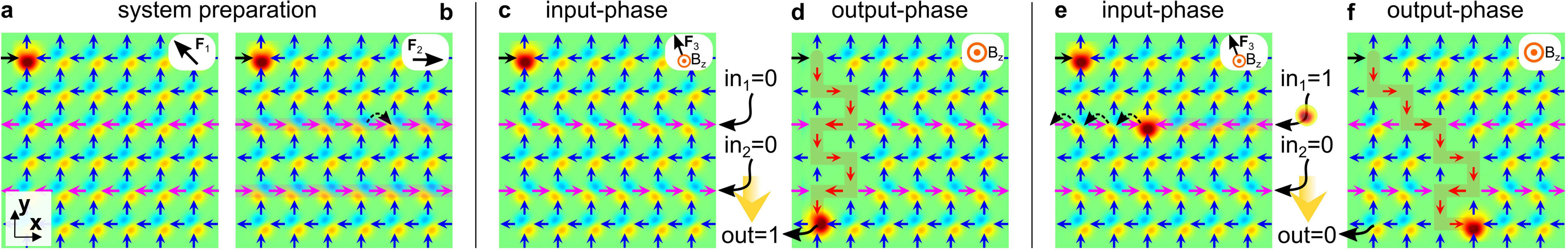}
\caption{Proposal of a NOR gate using the colloidal ice. 
(a,b) System preparation: a force $F_1$ creates a 
biased state (a), while another force $F_2<F_1$ 
reverses the spins only along two rows of 
particles that have smaller susceptibility (b).
(c,d) Images showing a $1$ output obtained from two $(0,0)$ inputs. The defect propagation is induced by 
a field $B$ applied perpendicular to the plane,
while an additional small force $F_3$ applied along the diagonal of the ice
lattice
prevents the
introduction of other defects from the upper left corner.
The final result shown in (d) is the $1$ output. (e,f) 
Images showing a $0$ output obtained from $(1,0)$ input.
In (e) the $1$ input of the first row containing
particles with high susceptibility is induced by an extra charge coming
from outside the region, from 
\textcite{Loehr2016b}.
}
\label{figure3}
\end{center}
\end{figure*}
One method for creating an artificial defect line is to start
with the square ice GS
(all type III vertices)
and flip particles in a sequence
that produces a double line of type IV vertices connecting two high energy type II and V
vertices. 
These defects can be described within the framework of the ``dumbbell" model \cite{Castelnovo2008a}. To each vertex one can associate an effective topological ``charge" $q$ that quantifies the degree of violation of the ice rule. As explained in Section II.1, a spin pointing towards (away from) the vertex center will have a positive (negative) charge $q_i$, and the total charge at each vertex $i$ will be $q=\sum_i q_i$. Thus, the GS of the square lattice corresponds to $q=0$ (type III), whereas type II (V) vertices have a positive $q=+2$  (negative $q=-2$) charge, Fig.~\ref{figure2}(a).
Using this formalism, it was shown that  
in a 3D spin ice and at low temperature,
topological defects interact only via a magnetic Coulomb law, $V(r) \propto Q^2/r$,
where $Q$ is the topological charge and $r$ is
the separation distance between the two monopoles.
However,
numerical simulations demonstrated that for a
2D square ASI, such strings shrink due to an additional line tension term \cite{Mol2009} that results from the missing degeneracy of this lattice at the single vertex level.
In the 2D case, the interaction potential between two defects becomes $V(l)=-Q/l+\upsilon l+c$, where $\upsilon$ is the
line tension and $c$ is a constant associated with the creation of defect pairs~\cite{Silva2013}.
Direct measurements of the average defect line length $\langle l \rangle$  showed that the defects shrink following overdamped dynamics described by 
\begin{equation}
\gamma \frac{dl}{dt} = -\frac{\partial V}{\partial l} = -\frac{Q}{l^2}-\upsilon \, \, , 
\label{eqchini}
\end{equation}
where $\gamma$ is a friction coefficient.
Eq.~\ref{eqchini} lacks the inertial term that is present when describing the motion of topological defects 
in ASIs~\cite{Vedmedenko2016}.
This first order nonlinear equation 
admits as solution the implicit function,
\begin{equation}
t-t_0=\frac{1}{\beta} \Big[ l_0-l+\sqrt{\alpha}  \Big( \arctan{\Big( \frac{r}{\sqrt{\alpha}} \Big) } - \arctan{\Big( \frac{l_0}{\sqrt{\alpha}} \Big) } \Big) \Big]  \, \, ,
\end{equation}
where $\alpha = Q/\upsilon$ is the ratio between the Coulombic and line tension contributions. 
The solution describes a monotonous shrinkage of the defect line starting from  $l_0$ at $t_0$, and was used 
to fit both the experiments and the simulation data. 
Such analysis
makes it possible
to quantify $\alpha=0.0290\pm 0.0014a^2$, where $a=29 \rm{\mu m}$
is the lattice constant of the square system.
This Coulombic charge is one order of magnitude lower
than the value calculated for magnetic ASI~\cite{Mol2009},
indicating that the line tension contribution is enhanced in the colloidal
ice.
Nevertheless, two topological charges
that are bound by a defect line in either
ASIs or in the colloidal ice interact with a similar Coulombic law. 

The modeling of the magnetic colloids is
similar to that used for charged colloidal systems
since the colloids are confined in double well traps and
experience a repulsive pairwise interaction with each other.
The colloid dynamics is governed by an overdamped equation of motion
given by
\begin{equation}
\frac{1}{\mu}
\frac{ d{\bf r}_{i}}{dt}= \sqrt{\frac{2}{D\,  dt}}k_B T N[0,1] + {\bf F}_{pp}^i + {\bf F}_s^i
\end{equation}
where $D$ is the diffusion constant, $\mu$ is the mobility,
$dt$ the time step, and $N[0,1]$ 
is a Gaussian distributed random number
with a mean of zero and a variance of 1. The first 
term is the thermal force. 
Magnetization of the colloids in the $z$ direction produces
a repulsive particle-particle
interaction force ${\bf F}_{pp}(r)=A_c{\bf \hat r}/r^4$ with $A_c=3\times 10^6\chi^2 V^2 B^2/(\pi\mu_0)$
for colloids at a distance $r$ apart,
where $V=\pi d^3/6$ is the colloid volume. 
The substrate force ${\bf F}^{i}_{s}$ traps the colloids in the double wells.
Simulations of magnetic colloids have focused on a range of parameters
that matches what is used in the experiments.

Since it is possible to tune the interaction forces between the
particles by changing the applied field,
it is interesting to model
the interaction between different types of monopoles
in a square ice for increasing field amplitude.
In the work of \textcite{Loehr2016b}
the magnetic colloids in a square ice substrate were initialized
in a GS containing two monopole excitations of opposite
charge and separated by a distance $d$.  These monopoles are attracted and approach each other at a velocity that increases
with the applied magnetic field.
At  high fields,
the monopoles annihilate more rapidly through a nucleation process in which
additional monopoles appear and break the defect string which extends
between the original two monopoles.
This work also showed that different monopole species
move at different velocities, in contrast to what would
be expected for magnetic spin ice systems.
By applying a periodic biasing field, the asymmetry in the monopole mobilities can be exploited in order
to create a ratcheting
motion of a defect line \cite{libal2017}.

It is also interesting to investigate grain boundaries (GBs) 
in the colloidal ice. In general, GBs are ubiquitous in condensed matter and they influence the performance of a variety of systems including high T$_c$ superconductors \cite{Graser2010}, organic films \cite{Rivnay2009}, and direct-band gap semiconductors \cite{Zande2013}. In the square colloidal ice, as in ASIs, GBs emerge as defect lines that separate unmatched regions of GS,  Fig.~\ref{figure2}(b). While GBs have been observed in magnetic samples \cite{Morgan2011,Zhang2013}, monitoring their formation and dynamics remains a challenging task due to the small length scales of the ASIs. In terms of plaquettes and not vertices, the GS of the square ice is composed of a checkboard pattern of
loops
with alternating chirality,
indicated by yellow and orange in the left panel of Fig.~\ref{figure2}(b). 
For the square ice this GS is twofold degenerate, and a 90 degree
rotation around a vertex produces another GS.
A domain wall, shown in the right panel of Fig.~\ref{figure2}(b), can be constructed by a line of achiral cells  that separate two incompatible regions of GS. These are very stable defects because, in contrast to a double line of 
type IV vertices,
it is necessary to modify all the vertices on one side of the domain wall to make them compatible with the GS of the other side.
This feature could be used to store binary information. 
By assigning a value of $0$ ($1$) to a clockwise
(counterclockwise) chiral cell,
it is possible to
use a binary representation to write
information in the GS of a lattice of colloidal ice or nanoscale ASI,
and later erase the information using a strong field which restores the full GS~\cite{Ortiz-Ambriz2016}.

A major interest in nanoscale magnetism and ASIs is the realization of logic circuits
based on a dense array of strongly magnetized elements. The particle-based ice could provide a guideline for the realization of simple and resettable logic ports based on magnetic dipolar interactions. Figure \ref{figure3} shows a proposal for a ``NOR" gate, a
functionally complete port capable of generating all logical functions~\cite{Bar91}.
The gate
is based on a
metastable biased state, filled by 
high energy type IV vertices, and
it can be initialized 
and reset by an external force $\bm{F}\sim (\bm{B}\cdot \nabla \bm{B})$ applied along one diagonal, $\bm{F}_1=F_1(\hat{\bm{y}}-\hat{\bm{x}})$.
In this system, the colloids in 
two horizontal rows of traps
have been replaced with particles that have
a higher magnetic susceptibility  (pink arrows in Fig.~\ref{figure3}).
A single particle in the upper left corner is
used
to trigger the propagation of a defect line. Then a second biasing force $F_2$ is
applied to flip only the two chains
containing high susceptibility particles. 
The input of the gate is
applied by switching the line of colloids with higher susceptibility, such that a flipped line means \emph{true} ($1$) and an unflipped line means \emph{false} ($0$). As shown in Fig. \ref{figure3}(c-d), the defect starts to propagate
under a perpendicular field $\bm{B}=B\hat{\bm{z}}$
and is deflected by the flipped input lines. After propagation, the location of the charge in the bottom row gives the output. If it is located in the lower left corner,
the output is ($1$). If either of the two input lines are in a $1$ state,
the defect line will not be deflected,
and the topological charge will end up in a different position,
giving an output of ($0$). 

\subsection*{5. Effect of disorder, doping, and system memory.}

A general open question in frustrated systems
is how robust or fragile the frustrated states are in the presence of
quenched disorder.
Numerical simulations of spatially extended samples with significant
statistics
can
shed light on this aspect.
In the colloidal artificial ice,
one method of introducing
quenched disorder is by randomizing the energy of the barrier at the
center of each double well.
In a square ice geometry subjected to a biasing field,
such randomness induces the formation of
$+1$ or $-1$ monopoles in the background of
biased ice rule obeying vertices.
\textcite{Libal2012a} studied the
effect of quenched disorder
on a colloidal square artificial ice
and found that under simulated annealing, instead of a completely
ordered state, the system formed
a partially disordered configuration containing
monopoles and biased ice rule obeying vertices.

\begin{figure}[t]
\includegraphics[width=\columnwidth]{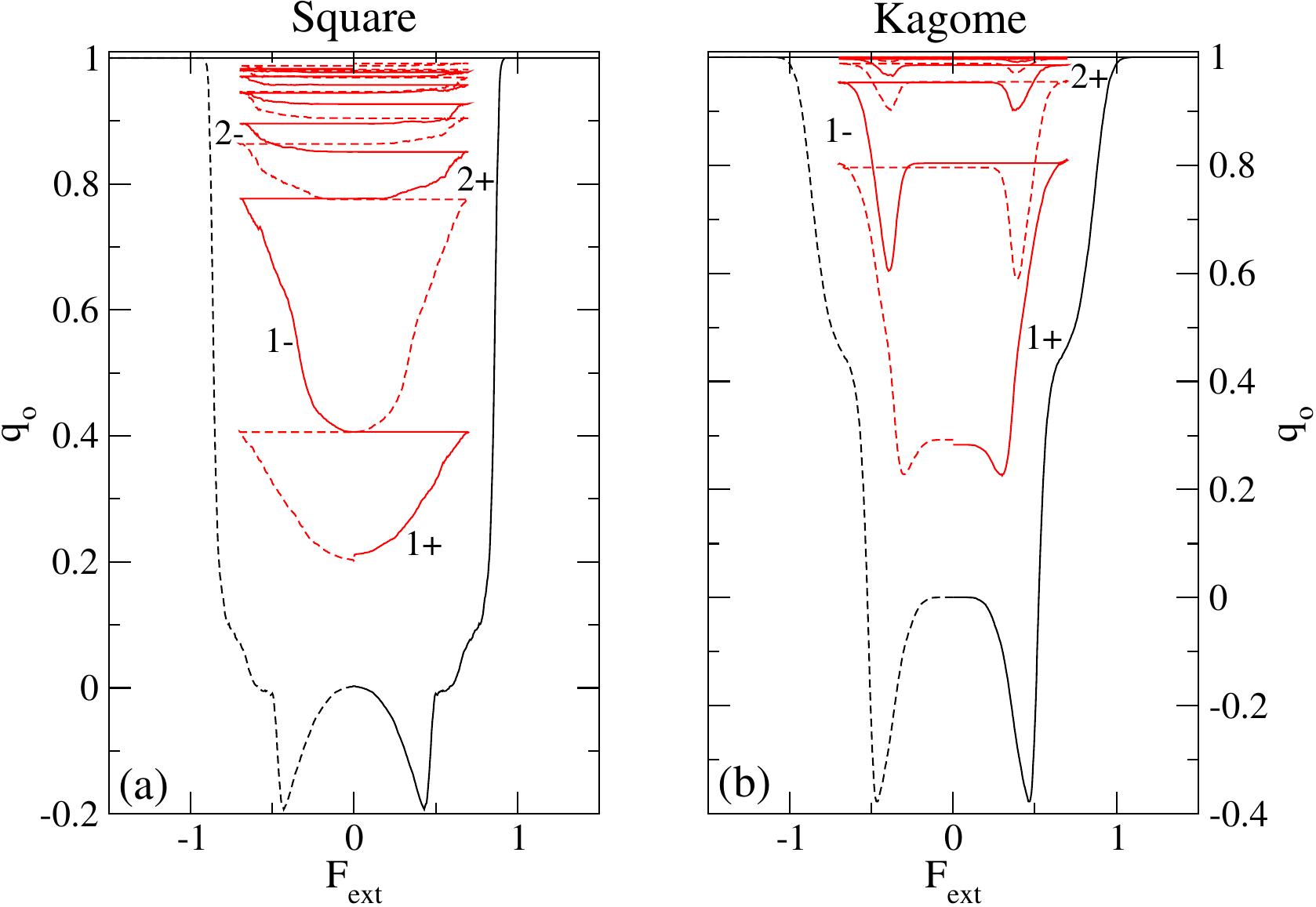}
\caption{Effective spin overlap $q_o$ vs biasing field $F_{\rm ext}$
  during consecutive hysteresis loops averaged over ten disorder realizations
  for (a) a square ice sample with disorder strength $\sigma=0.1$
  and (b) a kagom{\' e} ice sample with $\sigma=0.1$.
  Outer line: Saturated loop with $F_{\rm max}=2.0$, including the initial curve. Inner lines:
  Unsaturated loops with $F_{\rm max}=0.7$, with $n$ increasing from bottom to top;
  the first few half loops are labeled.
  Solid lines: clockwise loops; dashed lines: counterclockwise loops.
  In the kagom{\' e} ice, $q_o$ approaches 1 after only a few cycles,
  while a much larger number of cycles are required before
  $q_o \approx 1$ in the square ice, from \textcite{Libal2012a}. 
}
\label{fig:3}
\end{figure}

Under the application of a cyclic biasing drive, the monopole defects begin to annihilate
until the sample reaches a reversible state in which
the same effective spin configuration appears during each biasing cycle,
so that the system exhibits what is known as
return point memory \cite{Pierce2005}.  
\textcite{Libal2012a} also showed that the return point memory
state is 
reached much faster in a kagom{\' e} colloidal ice than in a square colloidal ice.
In the kagom{\' e} ice, irreversibility is produced by the motion of individual defects,
which can be pinned rapidly by the quenched disorder,
whereas in the square ice,
the irreversible motion arises from the reconfiguration of grain boundaries, and these
extended objects require a longer time to find a pinned configuration.

\begin{figure}
\includegraphics[width=\columnwidth]{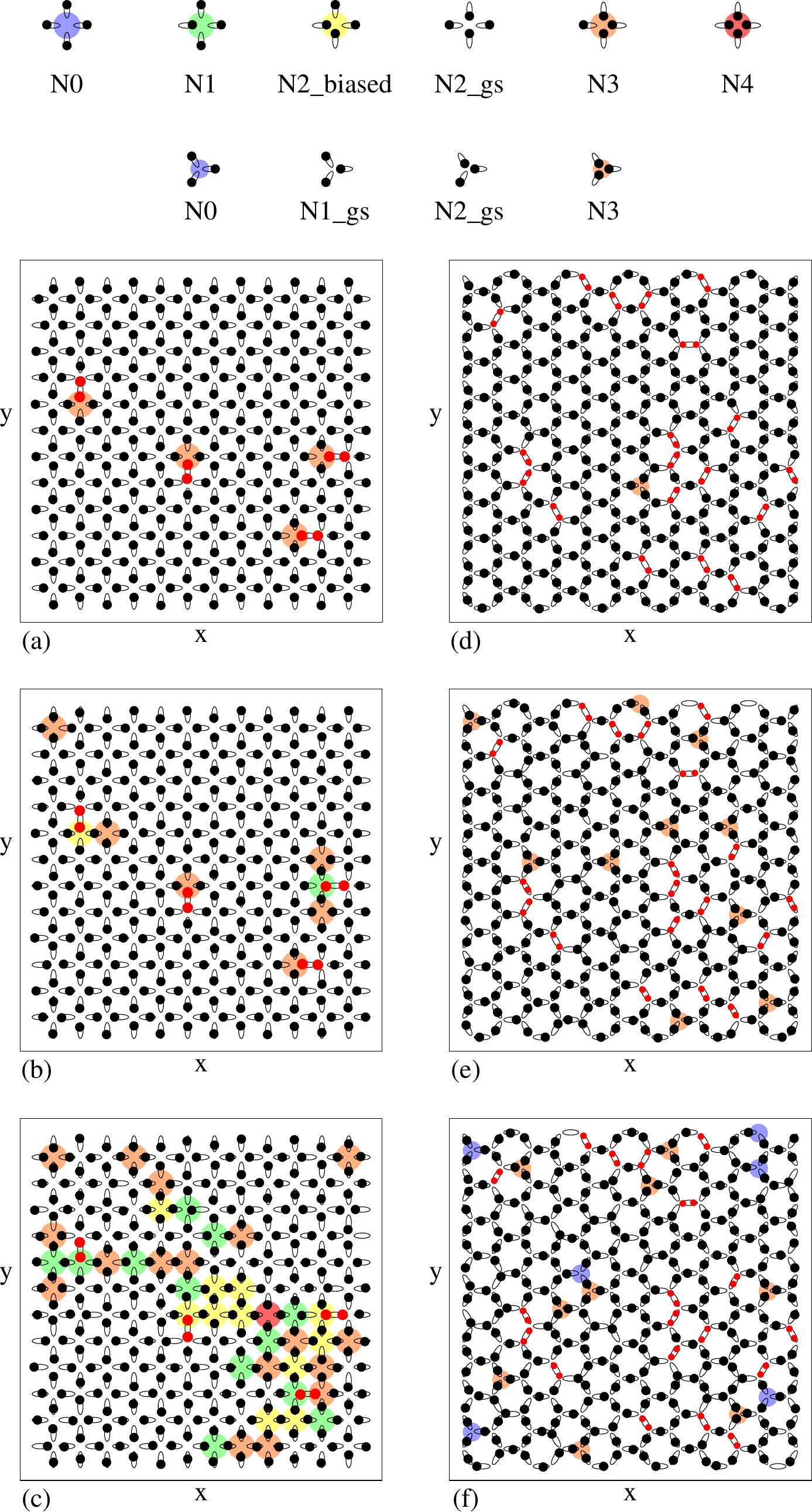}
\caption{Small portions of the square (left) and kagom{\' e}
(right) colloidal ice with double-well traps (open ellipses),
colloids in singly-occupied traps (filled black circles),
  and colloids in doubly-occupied traps (filled red circles).
  The larger colored circles indicate the vertex type as illustrated at the
  top of the figure: for the square ice,
  N0 (blue), N1 (green), N${2}_{{\rm{biased}}}$ (yellow),
  ground state N${2}_{{\rm{gs}}}$ (white), N3 (orange),
  and N4 (red); for the kagom{\' e} ice, N0 (blue),
  ground states N${1}_{{\rm{gs}}}$ and N${2}_{{\rm{gs}}}$ (white),
  and N3 (orange). (a), (b), (c)
  The square ice system at a doping of $\sigma =0.0236.$
  (a) At $T=1.0,$ below melting, each doped site is screened by an
  N3 monopole. (b) At $T=5.0,$ some thermal wandering of the N3
  sites can occur, and N1 states can form at the doped sites.
  (c) At T = 12 the regions away from the doped sites remain in the
  ground state while local melting occurs at and near the doped sites.
  (e), (f), (g) The kagom{\' e} ice system at a doping of $x = 0.095$.
  (e) At $T=1.0,$ below melting, the ground state absorbs the doping
  charge without forming defects by increasing the ratio of
  N${2}_{{\rm{gs}}}$ to N${1}_{{\rm{gs}}}$ vertices.
  (f) At $T=12.0,$ N3 monopoles form in regions away from the doped sites.
  (g) At $T=15.0,$ N1 monopoles begin to appear, from \textcite{Libal2015}.
}
\label{fig:4}
\end{figure}

Return point memory can be quantified by examining the overlap function $q_o$
which compares the effective spin 
configurations from one cycle to the next at the same biasing field, 
\begin{equation}
q_o(F_{\rm ext}) = N^{-1}\sum^{N}_{i=1} S^{n-1}_{i}(F_{\rm ext})S^{(n)}(F_{\rm ext}).
\end{equation} 
Here $S_i$ is an effective spin equal to $S_i=1$ ($S_i=-1$) if the colloid is sitting in the right or top end of the trap (left or bottom end), and $n$ is the number of cycles that the external force $F_{\rm ext}$ is applied.
When $q_o=1.0$, the spin configuration is perfectly reversible and is exactly the
same during each biasing field cycle,
while $q_o=-1.0$ would indicate that the spin configuration is exactly the opposite of its
previous value.
Fig.~\ref{fig:3}(a,b) shows $q_o$ versus
$F_{\rm ext}$ for the square and the kagom{\' e} colloidal
artificial ices
containing random disorder in the barrier heights implemented using a
Gaussian distribution of width $\sigma$.
The kagom{\' e} system is characterized by a modified version of the ice rule with 2-in/1-out ot 1-in/2-out vertex types. 
With repeated cycles of the biasing field,
$q_o$ approaches $1.0$, indicating the emergence
of return point memory.  This process takes about $n=10$ cycles
in the square ice but only $n=4$ cycles in the kagom{\' e} ice.
The steady state arrangement of artificial spins is not ordered
but contains numerous defects; however, the same pattern of defects
appears after each biasing field cycle.
Return point memory effects have also been observed in ASI by \textcite{Gilbert2015a}.
It would be interesting to understand how these
effects appear in many of the other
proposed ASI geometries \cite{Morrison2013}.
Further modeling work on kagom{\' e} and square
colloidal artificial ices that took into account the effects of temperature showed that
when biasing fields are present 
or when the system is prepared in different metastable states,
a series of structural transitions can occur, and that states with
monopole ordering appear \cite{OlsonReichhardt2012}.

In ASI,
each spin is constrained to point along the axis of the magnetic
nanoisland,
although is possible 
to remove entire islands.
In particle based artificial ices, it is possible to
introduce doubly occupied sites,
which would be like having spins pointing in both directions on
the same nanoisland.
Conversely, it is also possible to remove a particle
to create a vacancy or dilution of the spin arrangement.
\textcite{Libal2015}
considered the square and kagom{\' e} geometries containing doubly occupied wells where the system was
initiated in a GS configuration and then subjected to an
increasing temperature.
The effects of the doping are very different in the two ice geometries.
Figure~\ref{fig:4} illustrates the effects of doping 
on square ice (left panels)
and
kagom{\' e} ice (right panels) when the number of dopants is held
constant but the temperature is increased. 
For the square ice at low temperature,
N3 (three-in/one-out) vertices form around each doping site
in order to screen the doubly occupied wells.
As $T$ increases,
N1 defects and 
N2 biased defects begin to appear near the doping sites.
Thus in the square ice,
the doping sites act like
weak spots or nucleation points
that generate increased hopping
of the colloids in the surrounding sites.  
In contrast, the kagom{\' e} ice
GS can readily absorb the doubly occupied sites without
creating monopole states, as shown in Fig.~\ref{fig:4}(d) at
low temperature.
This is because the kagom{\' e} GS consists of an equal number
of N$1_{gs}$ and N$2_{gs}$ vertices, and the system can shift this balance
so that there are slightly more N$2_{gs}$ vertices, enabling it to 
incorporate the extra charge of the doubly occupied sites while still
remaining in a GS configuration. 
As the temperature increases, the doped sites in the kagom{\' e} lattice
reduce rather than increase the amount of hopping occurring in the
neighboring vertices.
\begin{figure}
\includegraphics[width=\columnwidth]{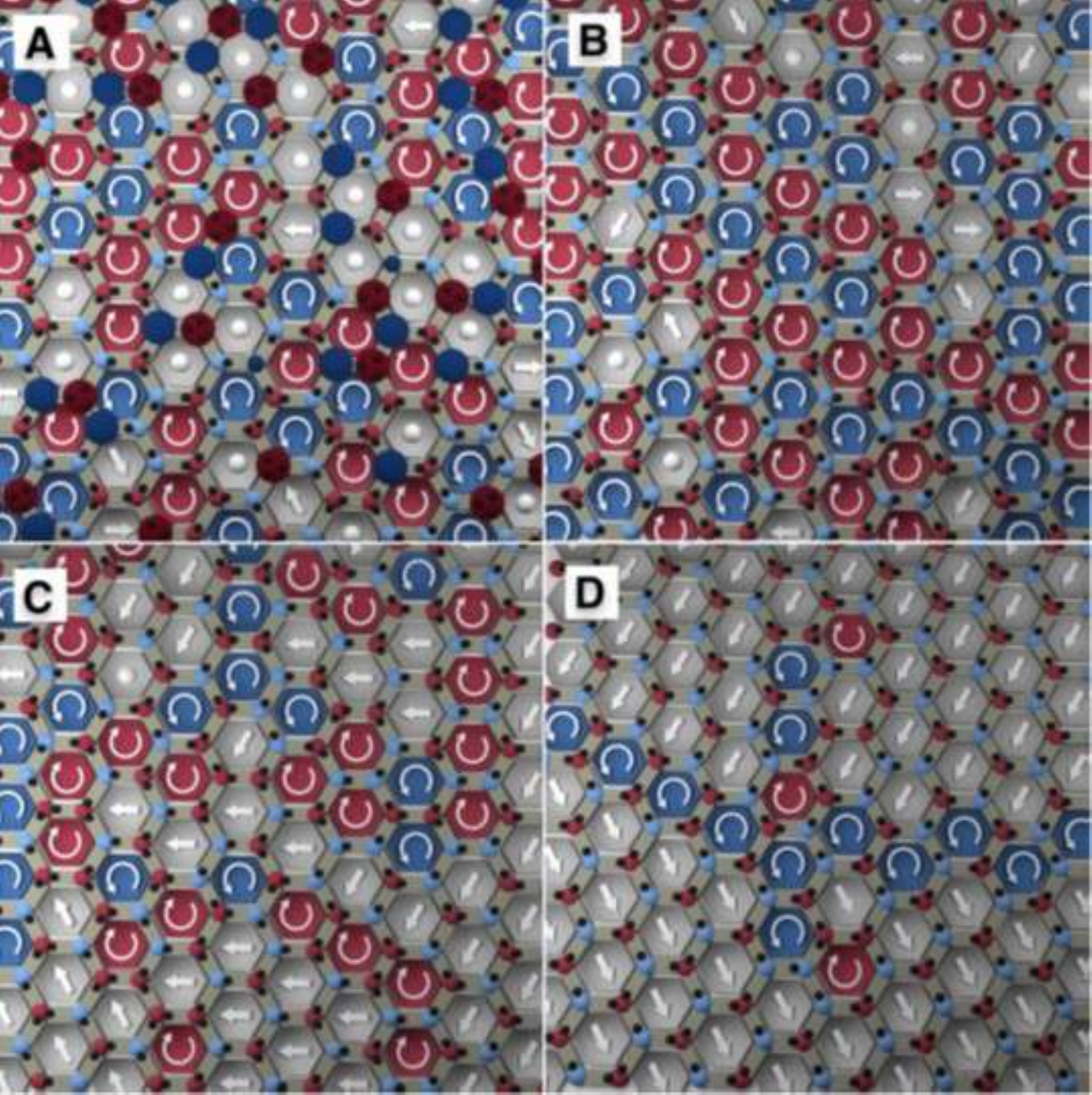}
\caption{Results from numerical simulations of the kagom{\' e} 
particle-based ice. Each double well trap (light gray) holds a single paramagnetic colloid (dark gray dots).  The hexagonal plaquettes contain arrows indicating the plaquette chirality direction or effective biasing
  field $F_b$ for chiral and achiral plaquettes, respectively.
  Dots indicate that no $F_b$ value can be assigned.
  (a) Paramagnetic phase at $B=0$ mT. Large red and blue disks indicate
  $q_n=\pm 3$
  vertices with
  $n=3$ and $n=0$, respectively.
  (b) Charge-free phase at $B=13.2$ mT containing only
  $q_n=\pm 1$
  vertices.
  (c) Partially charge ordered phase at
  $B=24$ mT with domains of charge and spin ordered vertices and plaquettes.
  (d) Ferromagnetic phase at
  $B=40$ mT containing a grain boundary.
  The system contains a second grain boundary with
  complementary chirality (not shown), from \textcite{Libal2018a}.
}
\label{fig:6}
\end{figure}
\subsection*{6.  Kagom{\' e} ice and its inner phase}

Kagom{\' e} systems have been the subject of intense research in nanomagnetism because this simple geometry provides a truly degenerate frustrated system \cite{Tanaka2006,Qi2008}.  The degeneracy arises at the single vertex level, where the three spins have equal distances and interaction energies. The ice rule of this lattice have associated a net topological charge, with $q=+1$ for $2$-in/$1$-out  and $q=-1$ for $1$-in/$2$-out. 
Although
the kagom{\' e} ice does not have a well defined GS, it was shown
theoretically in ASIs that the long range tail of
magnetic dipolar interactions can further reduce the entropy and
give rise to ordered phases. Specifically, multipole expansion \cite{Moller2009} and numerical simulations \cite{Chern2011} predicted that lowering the temperature should favor the transition from a short range ordered phase (``ice I") to a long range ordered state (``ice II"). In the latter case the spins on the hexagons define chiral and achiral loops along the lattice. Finally, the system
can transition into a completely ordered ``spin solid", or chiral phase, with chiral and achiral loops alternating in a chessboard-like pattern.
Direct visualization of these exotic states with their associated relaxation dynamics remains elusive. An indirect signature of the ice I phase has been reported
based on magnetotransport measurements of the Hall signal in a cooled ASI sample~\cite{Branford2012}. Further, the ice II phase was visualised via MFM on a lattice of permalloy nanoislands heated above the Curie temperature of the constituent material~\cite{Zhang2013}.  
More recently, low energy muon spectroscopy 
was used to probe
the existence of peaks in the muon relaxation rate that can be 
identified with a critical temperature 
associated with a phase transition that bridges such phases~\cite{Anghinolfi2015}.
Numerical simulations
of the magnetic colloidal system in the kagom{\' e} ice geometry show the emergence of different states 
as the colloid-colloid interaction strength
changes~\cite{Libal2018a}.
For weak interactions, the system is in
a paramagnetic state,
as shown in Fig.~\ref{fig:6}(a).
As the interaction strength increases, the system first enters
a state 
where the ice rule is obeyed in an ensemble of disordered spins,
as illustrated in Fig.~\ref{fig:6}(b).
This is followed by the appearance of the topologically ordered charged state
shown in Fig.~\ref{fig:6}(c),
and finally, in the limit of very strong interactions,
a three fold 
degenerate ferromagnetic state
emerges as shown in Fig.~\ref{fig:6}(d).
Unlike the phases predicted and experimentally observed with ASI, the colloidal kagom{\' e} ice forms a ferromagnetic state
at high interaction strength.
This ferromagnetic state
arises due to interactions between non-nearest-neighbors, rather than simple vertex energetics, and disappears
when the interactions are short ranged,
explaining why it was not  observed  previously.

\subsection*{7. Other geometries}

\begin{figure}
\includegraphics[width=\columnwidth]{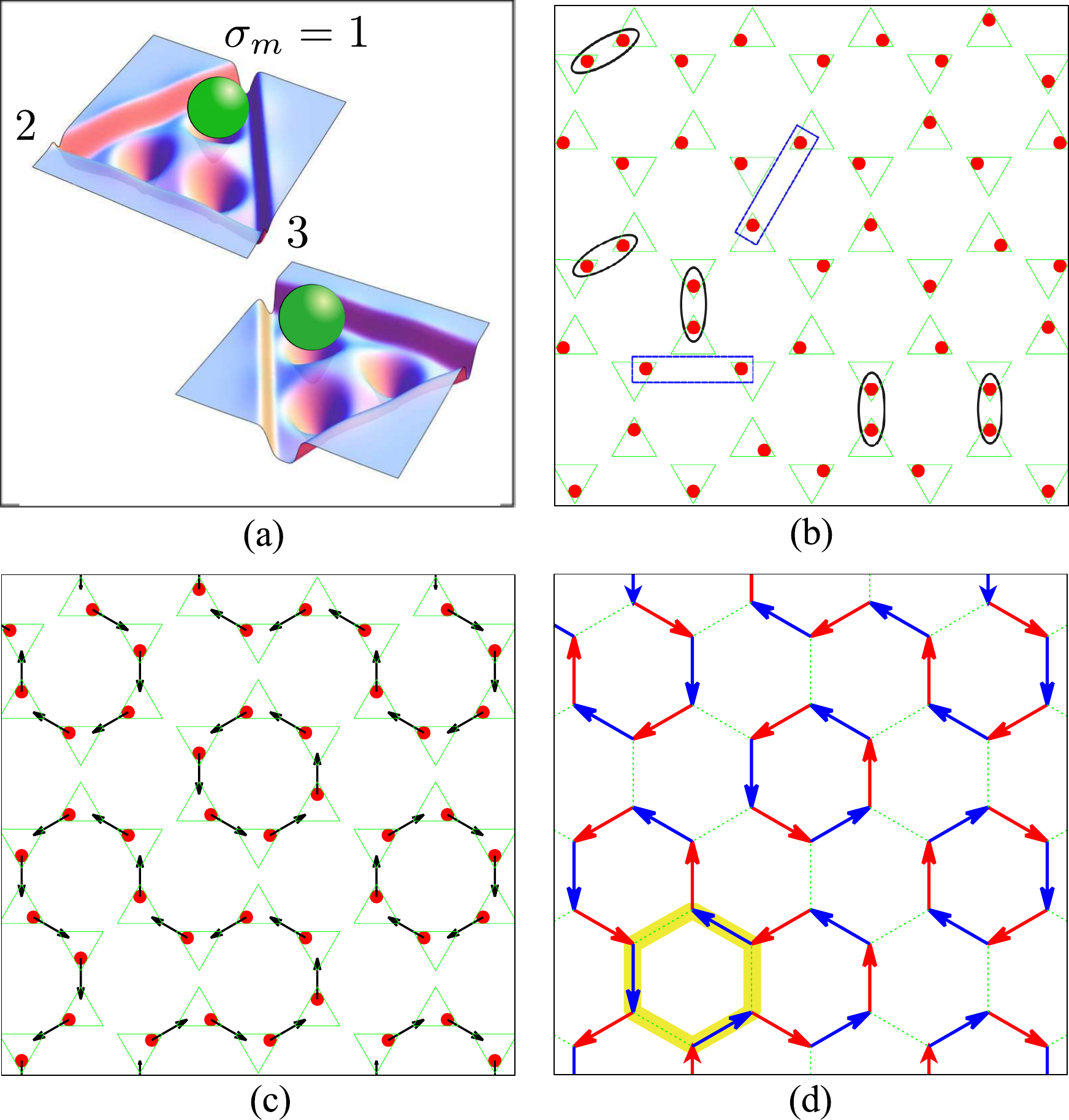}
\caption{ (a) Schematic diagram of the basic unit cell with two triple well traps, each containing one colloidal particle; (b) and (c) are snapshots of a small portion of the system. The green
triangles  represent  the  traps  and  the  red  dots  denote  the
particles.
(b) Random distribution of particles at high temperatures;
(c) shows an example of a particle configuration that can be mapped to
  random fully packed loops in the hexagonal lattice, as illustrated in (d).
  The broad yellow contour in (d) corresponds to a flippable type-II loop,
from \textcite{Chern2013}.} 
\label{fig:5}
\end{figure}

In addition to the square and kagom{\' e}
artificial ices, other types of frustrated systems have
been realized with microscale colloids.
\textcite{Chern2013} 
proposed a system of colloids
interacting with a honeycomb array of optical traps that 
contain three wells instead of two,
Fig.~\ref{fig:5}(a).
This system is a realization of a fully packed loop (FPL) model
and Baxter's three-coloring problem \cite{Baxter1970}. 
As a function of temperature and interaction strength,
a series of phases appear in the triple well system,
including a stripe state,
stripes with sliding symmetries, random packed loop states, 
and disordered states containing broken loops.
Figure~\ref{fig:5}(b) illustrates
the disordered state, while Fig.~\ref{fig:5}(c) and (d) show
the random FPL states. 

\begin{figure}[t]
\begin{center}
\includegraphics[width=\columnwidth,keepaspectratio]{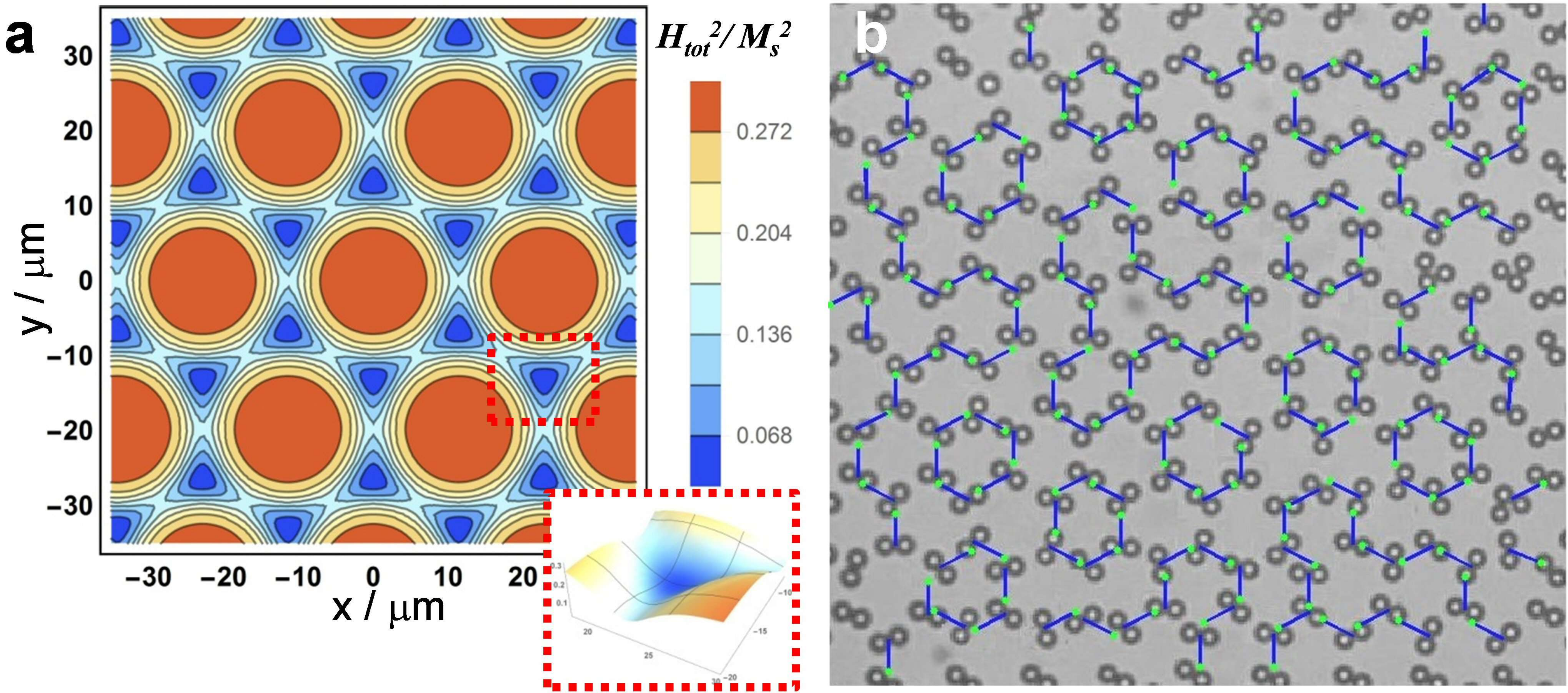}
\caption{(a) Magnetostatic energy 
landscape of a ferrite garnet film calculated under a static,
perpendicular magnetic field $B=3.6 \rm{mT}$. The inset shows a 3D
view of one triangular minimum. (b) Random FPL phase observed with a monolayer of paramagnetic colloids above the FGF after an annealing performed with a
precessing magnetic field, from~\textcite{Tierno2016b}.}
\label{Expfig5}
\end{center}
\end{figure}
FPL models
have been used to explain a broad class of phenomena in
magnetism, optics, and polymer physics \cite{Blote1994,Duplantier1998,Holleran2008,Jaubert2011}, but
physical realizations are rather scarce. 
An experimental realization of the colloidal version 
has been achieved
using a honeycomb lattice of triangular shaped magnetic minima,
as shown in Fig.~\ref{Expfig5}(a).
The complex magnetic potential arises when
a uniaxial ferrite garnet film (FGF)
is subjected to a constant magnetic field $\bm{B}_z=B\hat{\bm{z}}$.
The FGF is composed of a triangular lattice
of magnetic ``bubbles,'' {\it i.e.}, cylindrical ferromagnetic
domains that are uniformly magnetized \cite{Tier2007,Tier2009}.
Paramagnetic colloids dispersed above this
honeycomb magnetic lattice
self-assemble into interacting microscopic dimers. To anneal the lattice
into its minimum energy state, the system is subjected 
to a precessing magnetic field consisting of
a combination of the perpendicular field 
and an in-plane rotating field
$\bm{B}_{xy} \equiv B_{xy}[\cos{(\omega t)}\hat{\bm{x}}-\sin{(\omega t)}\hat{\bm{y}}]$. 
The resulting field $\bm{B}=\bm{B}_z+\bm{B}_{xy}$
performs a conical precession
around the $\hat{\bm{z}}$ axis with angular frequency $\omega$,
and sets the dimers into rotational motion. Depending
on the field parameters, the resulting dimer arrangement
can be mapped to a long range striped phase or to a random
FPL state, as illustrated in Fig.~\ref{Expfig5}(b). The mapping of the dimers
to the FPL model
is achieved by using the 
arrow representation
originally introduced by Elser and Zeng for the spin-$\frac{1}{2}$ kagom{\'e} antiferromagnet \cite{Elser1993}.
Since the dimer sits on one of the three sides
of a triangular minimum,
an arrow can be defined that points from the dimer center to the free corner of the
triangle.
This is shown by
a green dot with a blue line in Fig.~\ref{Expfig5}(b).
FPLs arise when arrows forming closed loops visit each
lattice vertex only once. 

Another recent realization of a particle ice composed by topographic double wells but in a different geometry 
is the highly coordinated triangular lattice ($z=6$) presented by \textcite{lee2018a}. 
Although not directly inspired by the water or spin ice materials, the triangular geometry reflects the spin disposition in several
magnetic materials with moments lying on weakly coupled parallel planes~\cite{Nakatsuji2005,Dublenych2017}. 
The triangular order reflects the natural arrangement of repulsive particles in the absence of the substrate. However, such ordering can be  frustrated by the presence of the central hill in the double wells.
In this geometry, collective interactions between the particles lead to a unique GS characterized by vertices with three colloids pointing inward and three outwards, similar to what was predicted for ASIs by~\textcite{Mol2012,Mol2013}. 
It was also found that the use of a bias force that magnetizes the system allows the GS to be accessed easily via a structural, martensitic-like transition characterized by the coherent sliding of one particle at each vertex.

\section{The fragile ice manifold}

\subsection*{1. Nature of the ice rule in colloidal ice}

While the ASI and the particle-based ice have often been considered as equivalent, their frustration and energetics are fundamentally different. At the nearest neighbor level,
a vertex in the particle-based ice has a lowest energy unfrustrated configuration in which
all particles are far from the vertex.
Indeed
the nearest neighbor energy of a magnetic spin ice vertex with
$n$ spins
pointing toward
the vertex is
proportional to the square of its topological charge,
\begin{equation}
E_n\propto q_n^2,
\end{equation}
thus favoring the ice rule, which minimizes the topological charge \footnote{Here we neglect, for the sake of argument, the known geometric
effects that can lift the degeneracy.}.
In contrast, the energy of a vertex in particle-based ice scales as
\begin{equation}
E_n \propto n(n-1),
\label{E_part}
\end{equation}
where $n(n-1)$ is the number of repulsive interactions among $n$ particles.
This energy favors  $n=0$ and $n=1$
states, which correspond to large negative charges
according to Eq.~(\ref{charge}).
In particular, the energies of the
particle-based ice
vertices do not posses a ${\cal Z}_2$ symmetry.
Thus the local energetics actually work against the ice rule.
The ice rule in particle-based ice
is only recovered in the thermodynamic limit and its origin is collective:
not all vertices can simultaneously have all particles located
away from the vertex.

In a finite size particle-based ice
system of uniform coordination,
the energy can be reduced by pushing particles onto the boundaries.
The total charge in the bulk is then bounded by the flow of the pseudo-spin field
through the boundary. Since the pseudo-spins have an absolute value of one,
the density of charge in the bulk scales at least as the reciprocal of the length of the boundaries, leading to an ice manifold in the thermodynamic limit.
In a system of multiple coordination, however,
such as $z=4,3$, we can consider the $z=3$ vertices as an ``internal boundary"
onto which the $z=4$ sub-lattice can push topological charges.
This would lead to the emergence of negative charges on $z=4$ vertices, in violation of the ice rule. Note that this behavior, which has been predicted~\cite{Nisoli2014,Nisoli2018a} and verified numerically and experimentally~\cite{Libal2018} points to an essential difference in the origin of the ice rule in magnetic and particle-based spin ices. The ice-rule in magnetic ice is known to be robust against decimation~\cite{Morrison2013},
mixed coordination~\cite{gilbert2014emergent, gilbert2016emergent}, and
 dislocations~\cite{drisko2017topological},
 and is present even in finite sized clusters~\cite{Li2010}. In particle-based ice, however, the local energy in Eq.(\ref{E_part}) {\it opposes} the ice rule, which is regained only as a collective compromise. Other differences appear in the kinetics.
For example, when defect lines in square colloidal ice are driven with a field,
the two monopoles at the ends of each line
have different mobilities since,
unlike the monopoles in magnetic spin ice, they have different energies~\cite{libal2017}.

\begin{figure}[t!!]
\begin{center}
\includegraphics[width=\columnwidth]{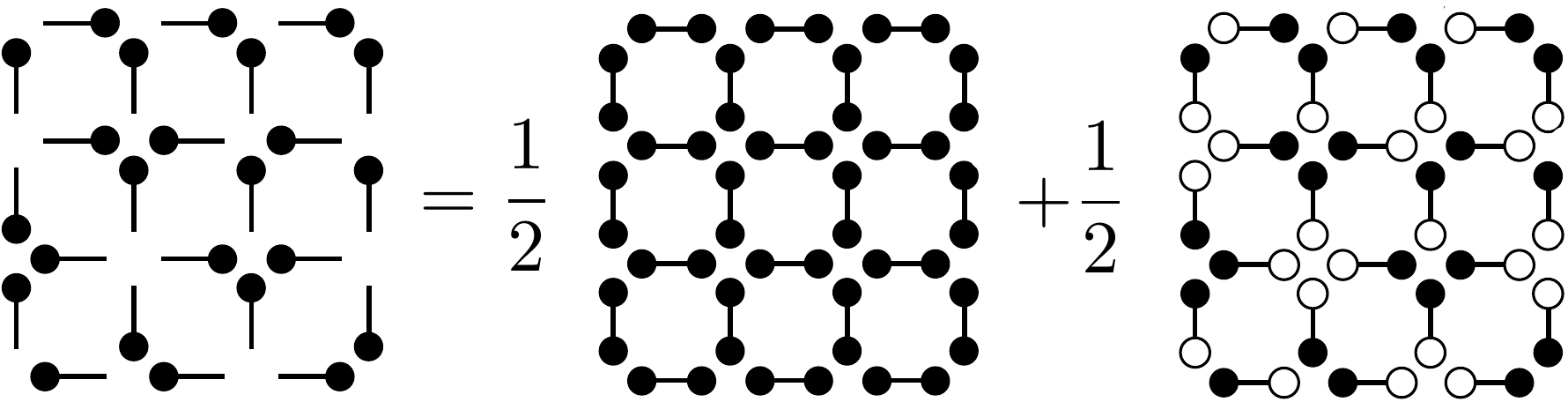}
\caption{Schematics showing that a colloidal ice, here in a random configuration, can be decomposed into a spin ice and a background exerting a  geometric field, from \textcite{Nisoli2018a}}
\label{fractionalization-a}
\end{center}
\end{figure}

The similarities and differences of the magnetic and particle-based ices can be quantified
exactly.
A mean field approach provides useful quantitative predictions
that are fit well by experimental and numerical results~\cite{Nisoli2014,libal2017}.
Starting with the approximation of
a  gas of decorrelated vertices,
a constraint must be introduced so that the total topological charge is conserved.
This takes the form of a Lagrange multiplier $\phi$ that modifies the original vertex energies from those in Eq.~(\ref{E_part}) to
\begin{equation}
\tilde E_n=E_n-q_n\phi.
\label{effE}
\end{equation}
For a lattice of coordination $z$,
the choice $\bar \phi \sim (z-1)$
ensures conservation of topological charge,
giving a spin-ice-like effective energetics,
$\tilde E_n\sim q_n^2$, and therefore a spin-ice behavior.
Eq.~(\ref{effE}) can be interpreted as follows: the {\it collective} effect of the particle-sharing vertices can be subsumed into a {\it emergent field} $\phi$, which modifies the energetics of the {\it individual} vertex. Indeed, in a better approximation, one can
allow the constraint-enforcing field $\phi(x)=\bar \phi + \eta(x)$ to fluctuate in space,
since it mediates an entropic interaction among the topological charges to which it is
coupled.
This results in a familiar Debye picture for purely entropic screening
with a correlation length $\xi^2 \sim T/ \overline{ Q^2}$ , where $\overline{ Q^2}$ is the charge fluctuation of the manifold.
For instance, in a fully
degenerate square ice,
$\overline{ Q^2}=0$
in the ice manifold
and thus the correlation length is infinite, as mentioned in the section on the topological properties of the ice rule.
In contrast,
$\overline{ Q^2} \ge1$
in kagom{\' e} ice
since each vertex has a charge of at least $\pm1$ and the ice phase is never critical. 

\begin{figure}[t!!]
\begin{center}
\includegraphics[width=.9\columnwidth]{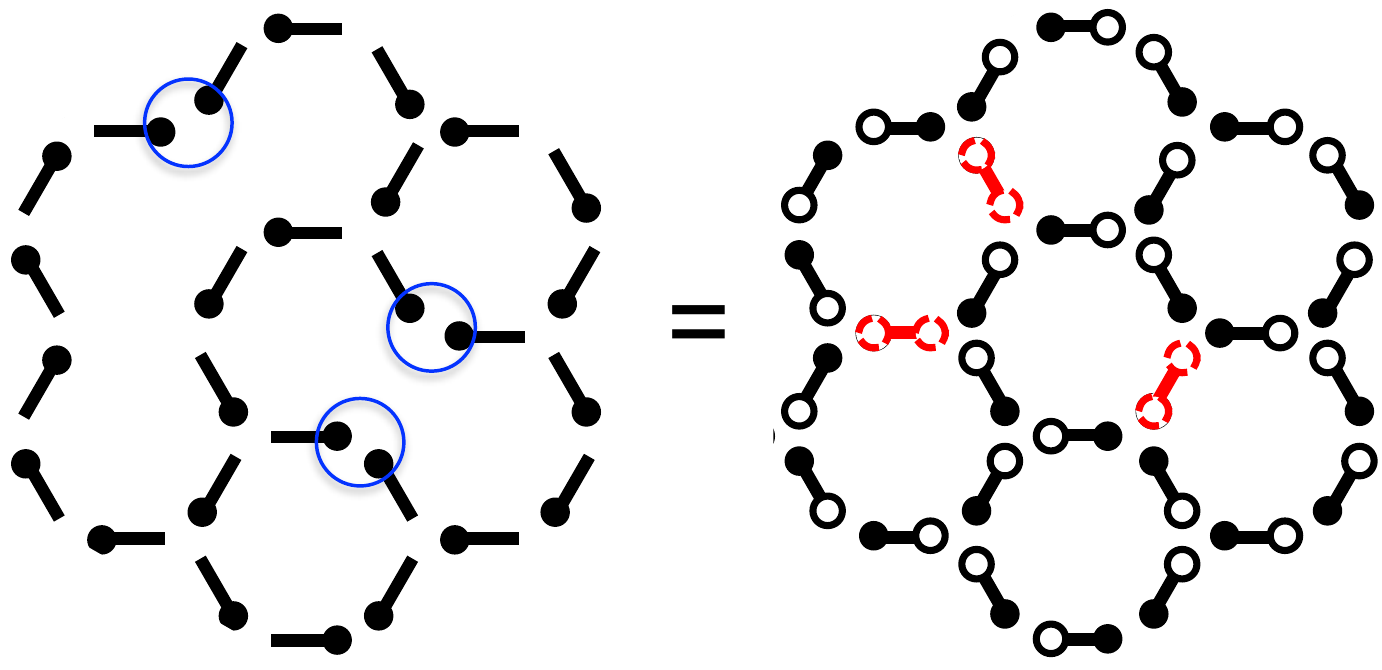}
\caption{Schematics showing that a particle-based ice obtained by removing traps from a regular lattice is equivalent to a spin ice stuffed with negative charges at the locations of the missing traps. This leads to violations of the ice rule through formation of positive topological charges (blue circles), from \textcite{Nisoli2018a} }
\label{fractionalization-b}
\end{center}
\end{figure}
Consider now a lattice of multiple coordination. It is impossible to find a value of $\phi$ that can  return a ${\cal Z}_2$ invariant, ice-like effective energetics in Eq.~(\ref{effE}) for both sublattices. One of the lattices must break the ice rule. 
How this happens can be understood by a geometrically intuitive, exact treatment~\cite{Nisoli2018a}. The energy of the system is given by 
\begin{equation}
H= \sum_{{\bf y}\ne{\bf y'}}\psi\left(|{\bf y}-{\bf y'}|\right)
\label{H0}
\end{equation}
where $\psi(r)$ is an isotropic repulsive interaction 
and ${\bf y}$ labels the position of the particles in the traps.
These positions represent a binary variable that we can represent as \textemdash\tikzcircle{3pt} or \tikzcircle{3pt}\textemdash. 
Indeed Eq.~\ref{H0} does not look like a spin ice Hamiltonian. 
Then we can ascribe a {\it positive} charge to the real  \tikzcircle{3pt} particles  and we can consider the empty locations  ${\bf y}^-$ of the traps as virtual {\it negative} charges \tikzcirclew{3pt}, which repel (attract) other negative (positive) charges. We can fractionalize a trap on an edge ${\bf x}$ as 
\begin{equation}
\operatorname{-----\tikzcircle{3pt}}=\frac{1}{2} \operatorname{ \tikzcircle{3pt}---\tikzcircle{3pt}}+\frac{1}{2} \operatorname{\tikzcirclew{3pt}---\tikzcircle{3pt}},
\label{iso}
\end{equation}
i.e. a  {\it positive dumbbell} $\operatorname{ \tikzcircle{3pt}---\tikzcircle{3pt}}$ (a  trap doubly occupied by positive charges) plus a {\it dipole} of negative and positive charges represented by a spin $\vec {\sigma} =  \operatorname{\tikzcirclew{3pt}---\tikzcircle{3pt}}$ located in ${\bf x}$, the center of the trap. Then the energy  in Eq.~(\ref{H0}) can    be rewritten  as
\begin{equation}
{ H}= \frac{1}{2} \sum_{{\bf x}\ne  { \bf x'}}{\sigma}^i_{\bf x} J_{ii'}\left({\bf x}-{\bf x'}\right)  {\sigma}^{i'}_{\bf x'}-\sum_{\bf x}\vec{\sigma}_{\bf x}\cdot \vec{B}({\bf x}).
\label{H}
\end{equation}
The first term is the interaction between dumbells and is clearly a spin-ice Hamiltonian. $J_{ii'}\left({\bf x}\right)$ is a tensor field and the background  field $\vec{B}$  mediates the interaction between dipoles and the positive dumbbells,
both of which can be reconstructed from the particle-particle repulsive interaction $\psi$. Fig.~\ref{fractionalization-a} illustrates this decomposition. 
It follows that a particle-based ice becomes equivalent to a magnetic spin ice when the second term in Eq.~(\ref{H}) is zero. That is certainly  true if a lattice has point reflection symmetry in the middle points $\{{\bf x}\}$ of each edge, and explains why  the
kagom{\' e} and square particle-based ices follow the ice rule, as found previously numerically and experimentally~\cite{Libal2006a,Ortiz-Ambriz2016,Loehr2016b}. 

\subsection*{2. Decimated systems}

\begin{figure}[t]
\begin{center}
\includegraphics[width=\columnwidth,keepaspectratio]{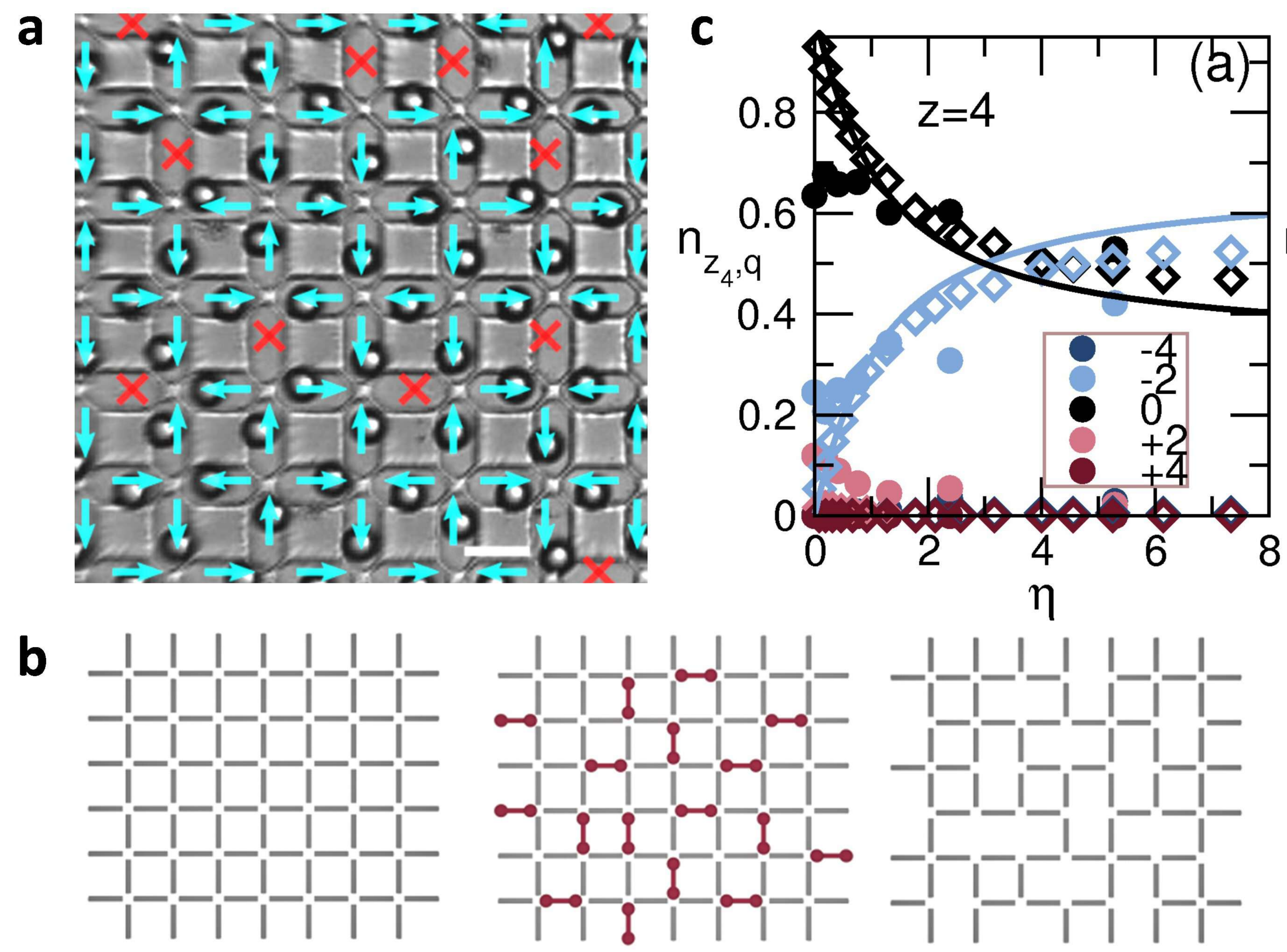}
\caption{(a) A decimated colloidal square ice 
where the missing spins (particles) are denoted by crossed arrows. (b) Decimation procedure for the square lattice that creates only vertices of coordination 
$z = 3$ and $z = 4$, but not $z = 2$. This process is equivalent to a partial dimer covering (red dumbbells) of the edges. (c) Fraction of vertices  $n_{z_{4,q}}$ vs decimation ratio  $\eta=N_{z_3}/N_{z_4}$ 
for vertices with coordination number $z = 4$ and at different 
value of topological charge $q$. Dark blue: $q = −4$; light blue: 
$q = −2$; black: $q = 0$; pink: $q = +2$; red: $q = +4$. 
The experimental results (bullets) and numerical results (diamonds) are compared to theoretical predictions (solid lines), adapted from \textcite{Libal2018}.}
\label{figure4}
\end{center}
\end{figure}

The analogy between the colloidal ice and nanoscale ASI
does not hold in general
for more complex geometries.
Imagine decimating
a simple lattice such as the kagom{\' e} lattice by removing some traps.
This breaks the reflection symmetry and the background field will perturb the spin ice energetics since the saturated dumbbells are replaced by
negatively saturated ones at the locations of the missing traps.
As shown in Fig.~(\ref{fractionalization-b}), the decimated system is equivalent to a spin ice stuffed with negative charges.
The charges polarize the dumbbells close to the decimated vertices, breaking the ice rule.
In these geometries with mixed coordination $z$,
the conservation of the topological charge holds only at the global level, not at the sub-lattice level.

In the case shown in Fig.~\ref{figure4}(a), the square system  ($z=4$) is decimated with optical tweezers to a lattice of mixed coordination, $z = 3,4$.
The decimation process
is equivalent to a partial, random dimer covering of the edges,
as illustrated in Fig.~\ref{figure4}(b). Randomly chosen edges of the square lattice are covered by dimers under the constraint
that each vertex contains at most one dimer. Removing an edge between two “dimerized” vertices with $z = 4$ gives rise to only vertices of $z = 3$. The process avoids formation of $z = 2$ vertices, and thus simplifies the vertex hierarchy of the mixed coordination system \cite{gilbert2014emergent}, although similar effects can also be observed in the presence of $z = 2$ vertices.  

Combined experiments and numerical simulations 
show that in such geometry,
the ice rules are spontaneously yet selectively violated via the formation of negative topological monopoles of charge $q = −2$ on the vertices of high coordination $z=4$.
The low coordinated vertices $z=3$
still obey the ice rule ($2$-in/$1$-out or $2$-out/$1$-in);
however, the relative ratio of $q = 1$ to $q = −1$ charges
changes in order to compensate
the negative charges that accumulate around the $z = 4$ vertices.
Even in this situation, the total topological charge of a system of ``dipoles" remains zero.

A quantitative analysis of the experimental and theoretical results provides
additional insight into the decimation process.
Figure~\ref{figure4}(c) shows the relative fraction of $z=4$ vertices, $n_{z_{4,q}}$, versus the ratio between the two vertex coordinations $\eta=N_{z_3}/N_{z_4}$,
grouped by topological charge. The number of negative $q=-2$ charges generated around the $z=4$ vertices increases with increasing $\eta$,
quantifying the strength of the ice rule violation.
Above a critical decimation threshold,
the entire system disorders due to the spontaneous appearance of entropy-driven negative monopoles which induce topological charge transfer between the sublattices.
As a result, the colloidal ice has a ``fragile" low-energy manifold
that is produced by an energetic compromise between locally excited vertices.
This is in contrast to magnetic ASI, which are structurally ``robust'' ices. 
These observations prompt different exciting ideas.
Since ice rule fragility is associated with topological charge transfer among sub-lattices,
these new phenomena can be exploited for domain wall engineering,
in which membranes that are semipermeable to the topological charge of defects
are structurally designed.
These results also apply to lattices of different coordination or to
ASI systems of nonzero residual entropy. 

\begin{figure}[t]
\begin{center}
\includegraphics[width=\columnwidth,keepaspectratio]{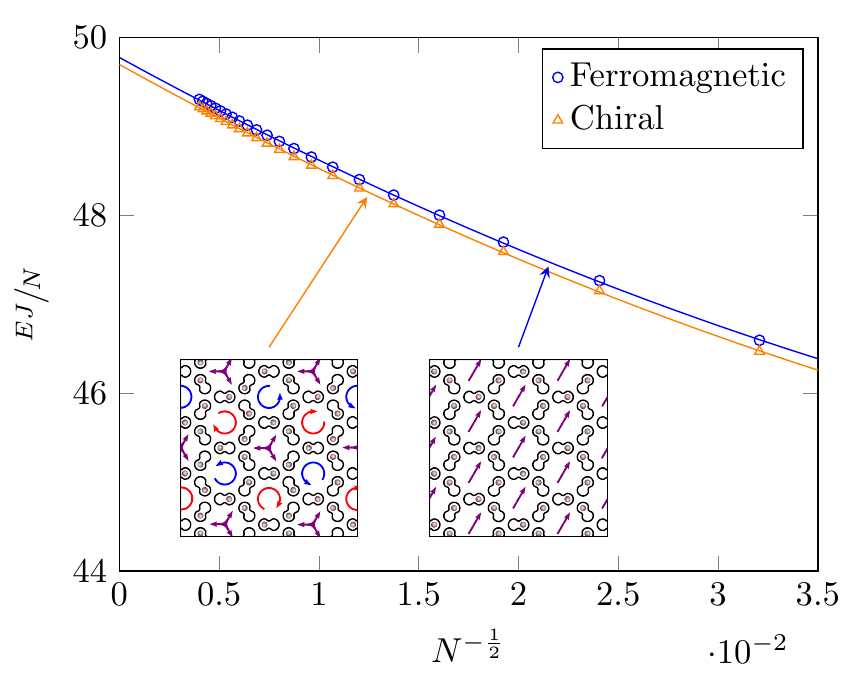}
\caption{Magnetostatic energy for $N = 3L^2$ interacting particles with
periodic boundary conditions for the chiral (red triangles) and ferromagnetic
(blue circles) phases of the kagom{\' e} ice (up to $L = 144$). Inset
shows schematics illustrating the two competing orders. Scattered
points are numerical calculations and continuous lines are polynomial
fits of the form $E/J \sim \alpha N + \beta N^{1/2} + \gamma$,
where $\alpha$ is the energy
per spin in the thermodynamics limit
and $\beta$ and $\gamma$ are finite-size corrections,
from~\textcite{LeCunuder2019}.}
\label{figure5}
\end{center}
\end{figure}

\subsection*{3. Discrete versus continuum models}
The interactions between the pseudospins in the colloidal ice are given by a generalized dipolar interaction constructed from the effective interaction $V(r)$ between pairs of colloids (which in the case of magnetic colloids is $V \sim r^{-3}$).
This implies that colloids with different repulsive interactions exhibit
different phases within their ice manifold,
and in particular,
that electrically charged colloids are expected to
more faithfully reproduce
the phases
of magnetic spin ices.  
Inspired by these results,
in a
theoretical study,
Le Cunuder {\it et al.} investigated the energetics and phase transitions that occur in the
kagom{\' e}
geometry by varying the temperature rather than increasing the interaction strength. Numerical calculations of the 
total magnetostatic energy for a system containing $N$ particles
showed that the chiral state (``spin solid") always has a lower energy than the ferromagnetic one, a result which is robust regardless of the 
system size, Fig.~\ref{figure4}(c).
This calculation was performed by assuming that the particles are fixed
at the bottom of each trap,
without considering any relative displacement. 

The apparent discrepancy between the analytic  results
and those obtained in the simulations and experiments described above
was resolved by
performing
two types of Monte Carlo simulations  \cite{LeCunuder2019}. In the first, discrete model, particles are only allowed to jump from
one side of the bistable traps to the other, while in the
second, continuous model, particles are allowed to make these same jumps
and also to move continuously around the lowest point of the bistable traps.
While the discrete simulations reproduced the well-known phase structure of the dipolar spin ice, the continuous Monte Carlo
simulations showed a single phase transition
from charge disorder directly to the ferromagnetic state. 
Furthermore,
using the continuous model, it was shown that by modifying the strength of the traps, the GS of the particle-based ice can be changed from the chiral ordering to the ferromagnetic one.

\subsection*{3. Future directions in particle ice}
The experimental realization of particle ice 
represents a starting point for exploring the novel physics that emerges from the mesoscopic character of the particles.  A variety of avenues
for future research are now available.
\\
\\
\textbf{Relaxation and dynamics:} Apart from the annihilation of simple defects,
the kinetic mechanisms behind the relaxation dynamics of the particle ice remain
largely unexplored.
Since
direct visualization of the dynamics is possible in
colloidal ice,
correlation functions for relatively long times can be extracted from
the particle positions,
and rearrangement events (spin flips) can be identified
as the system ages from a metastable state.
An interesting question is whether the colloidal ice
could exhibit glassy behavior and kinetic arrest. The presence of disorder in the system,
such as an inhomogeneous distribution of the hills within the double wells,
or induced decimation,
may make the energy landscape more rugged and induce glassy behavior.
There have been reports of
evidence of glassiness in frustrated systems from  both
theoretical models and experiments.
For example, a spin glass phase was predicted
to occur for a decimated square ice in an ASI \cite{Sen2015},
and experiments with buckled colloidal monolayers \cite{Zhou2017} 
showed evidence of dynamical arrest.
\\
\\
\textbf{Thermalization effects:}
Using smaller particles with large thermal fluctuations would allow
spontaneous particle switching within the double wells, a feature that is currently absent in the experimental system. This platform could be used to explore thermalisation effects in the colloidal ice.  Such an investigation could be performed by
using optical tweezers to prepare a square
or kagom{\' e} ice lattice in the lowest energy state while keeping the
particle-particle interaction strength high.
Subsequently
lowering the magnetic field
would then favor thermal disordering
starting from the GS and make it possible
to determine the lifetime, fraction, and dynamics of the
emerging, thermally induced ``Dirac'' strings.
\\
\\
\textbf{Complex geometries:}
The lithographic approach imposes no limit on the type of two-dimensional structures that can be engineered.  While 
most of the works on particle based artificial ices
have focused on square or 
kagom{\' e}  lattices, numerous other geometries  that have
been proposed for ASI \cite{Morrison2013}
could also be realized in particle based systems.
Since the
particle ice system minimizes
the global energy rather than the local vertex energy, it often exhibits
fragility \cite{Libal2018}, so 
it is likely that these systems
could have very different behaviors than their ASI counterparts.
In addition, aperiodic structures or even disordered lattices with hyperuniform
properties can easily be designed and implemented with the colloidal ice.\\
\\
\textbf{Annealing procedures:} Frustrated systems can easily be trapped in a metastable state when the material or sample is cooled to reach the GS. Different annealing protocols have been developed to date, such as thermalisation during sample growth \cite{Morgan2011} or above the Curie temperature of the constituent material \cite{Zhang2013},
and rotational demagnetization \cite{Wang2018,Nisoli2007,Ke2008}. However, most of the theoretically predicted ordered phases in highly frustrated systems are still far from being realised since these techniques do not allow for direct system visualization during annealing, making it impossible to monitor in situ how close or far from the GS the system is as a function of time. The colloidal ice could provide a way to directly follow the annealing process in-situ. The current realization relies on static fields; however, it would be possible to introduce a time dependent field that spins the particles in order to induce tunable anisotropic or even time-averaged attractive in-plane interactions. 
\\
\\
\textbf{Three dimensions:}  A major effort of the frustrated magnetism community is
devoted to the realization of a 3D artificial version of the pyrochlore lattice,
such as by using magnetic wires or nanobars oriented at determined angles
in order to replicate the tetragonal order \cite{Pacheco2017}.
Recently, an alternative approach to this goal has been proposed \cite{Mistonov2013} based on the realization of an inverted colloidal opal filled with cobalt via electrochemical crystallization. This opal was obtained using the colloidal crystal template technique, which is a well-established method in material science \cite{Zakhidov897}, based on replicating the long-range order of an assembled colloidal crystal in a solid matrix. Although this type of approach may restrict the lattice geometry \cite{chern2014realizing}, or lead to artifacts due to domain wall pinning within the magnetic network, it represents a rather simple, fast and versatile method of obtaining a highly ordered 3D porous structure. Future attempts at the realization of a 3D colloidal ice system may exploit similar self-assembly techniques. On the other hand, progress in optical manipulation has made it possible to create 3D optical traps for colloids \cite{Grier2006},
and thus
it should be possible to create new types of fully 3D spin ice
geometries using optically confined colloids. 
In principle, such geometries
could be used to model water ice more accurately,
or possibly as a method for realizing deconfined phases.\\

\section{Other particle based frustrated systems}
\begin{figure}
\includegraphics[width=\columnwidth]{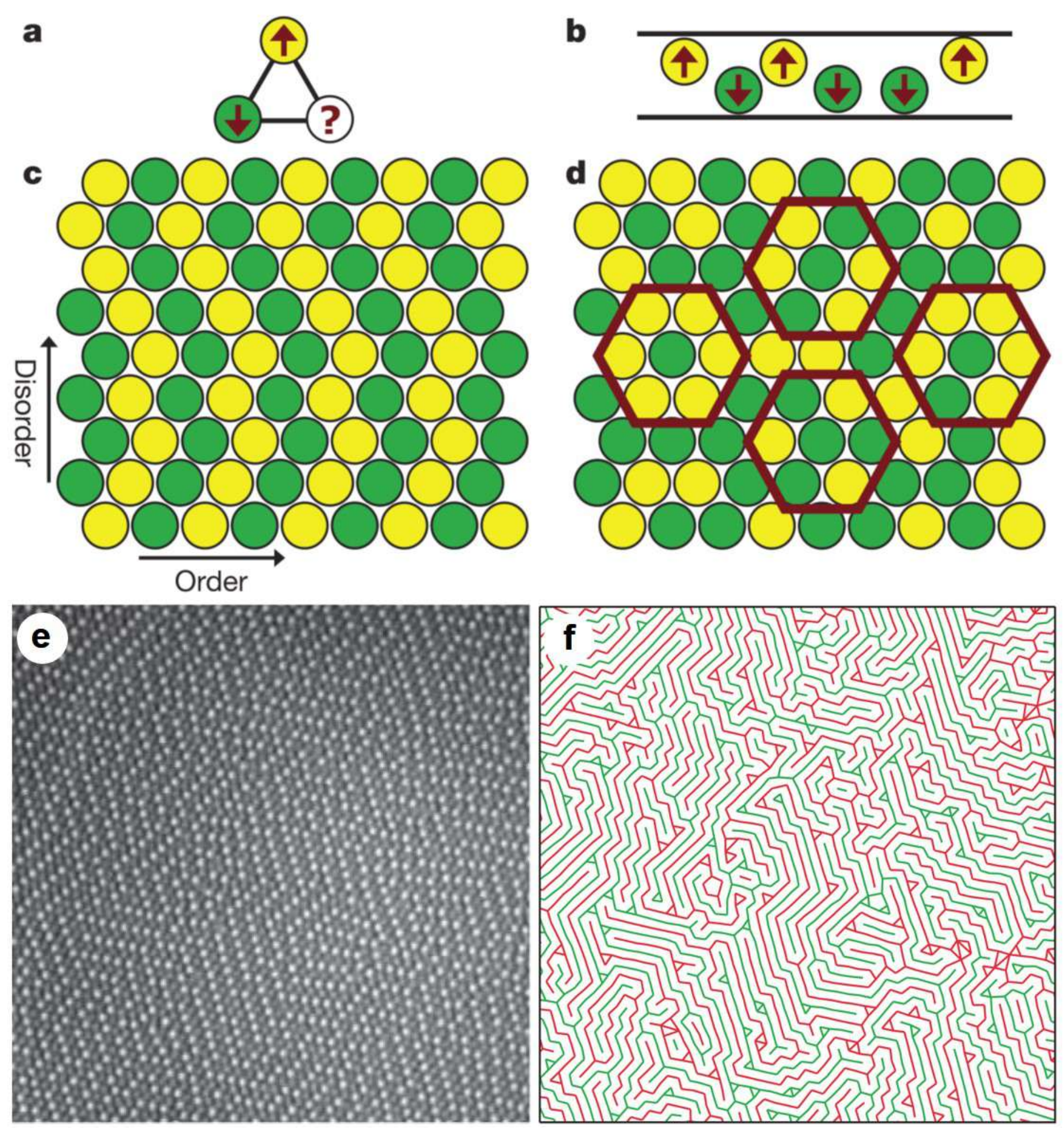}
\caption{(a) Three spins on a triangular plaquette cannot simultaneously satisfy all
  antiferromagnetic interactions. (b) For colloids confined between walls separated by a
  distance of the order 1.5 sphere diameters (side view), particles move to opposite walls
  in order to maximize their free volume. (c, d) Ising-like GS configurations in which
  each triangular plaquette has two satisfied bonds and one frustrated bond.
  (c) Zigzag stripes generated by stacking rows of alternating up/down particles with
  random sideways shifts, where all particles have exactly 2 frustrated neighbors.
  (d) Particles in disordered configurations have 0, 1, 2 or 3 frustrated neighbors
  (red hexagons).
(e) Microscope image of a buckled system where bright (dark) particles are close to the upper (lower) plane. (f) Corresponding labyrinth patterns with the frustrated up–up (down–down) bonds draw in red (green),
adapted with permission
from~\textcite{Han2008}}
\label{fig:7}
\end{figure}

Frustrated configurations have also been realized for colloidal 
systems in the absence of a substrate~\cite{Han2008}. 
For example, when colloids confined to a 2D plane are allowed to buckle into the
third dimension, as shown in Fig.~\ref{fig:7}, frustration emerges since upward
and downward buckling are equal energy states.
In the triangular lattice naturally formed by the colloids,
the buckling process produces a structure
similar to an antiferromagnetic Ising model on a triangular 
lattice \cite{Shokef2013},
and interesting disordered and stripelike patterns appear. While the original Ising antiferromagnet at low temperature features 
extensive entropy~\cite{wannier1950,Wannier1973}, the buckled system displays subextensive one, which points towards the presence of glassy dynamics and kinetic arrest. Indeed glassiness in such system has been recently reported~\cite{Zhou2017}, with other features as the emergence of an order by disorder transition \cite{Shokef2011,Leoni2017}.
Other colloidal systems that can exhibit frustration effects
include either particles with complex shapes~\cite{Brown2000}, or isotropic ones deposited above 
deformed surfaces~\cite{Bausch2003,Soni2018}. In the first case, the frustration
can be tuned by designing different shapes which compete with the confinement and impede crystallization, often producing a disordered state~\cite{Zhao2009,Zhao2015}. 
Tuning the shape of the confining surface may also lead to 
frustration effects between isotropically repulsive colloids. The surface topology will induce the formation of lattice defects as disclinations, releasing energetic stresses arising form the packing on the curved confinement~\cite{Nelson2002,Irvine2012,Guerra2018}.

\begin{figure}
\includegraphics[width=\columnwidth]{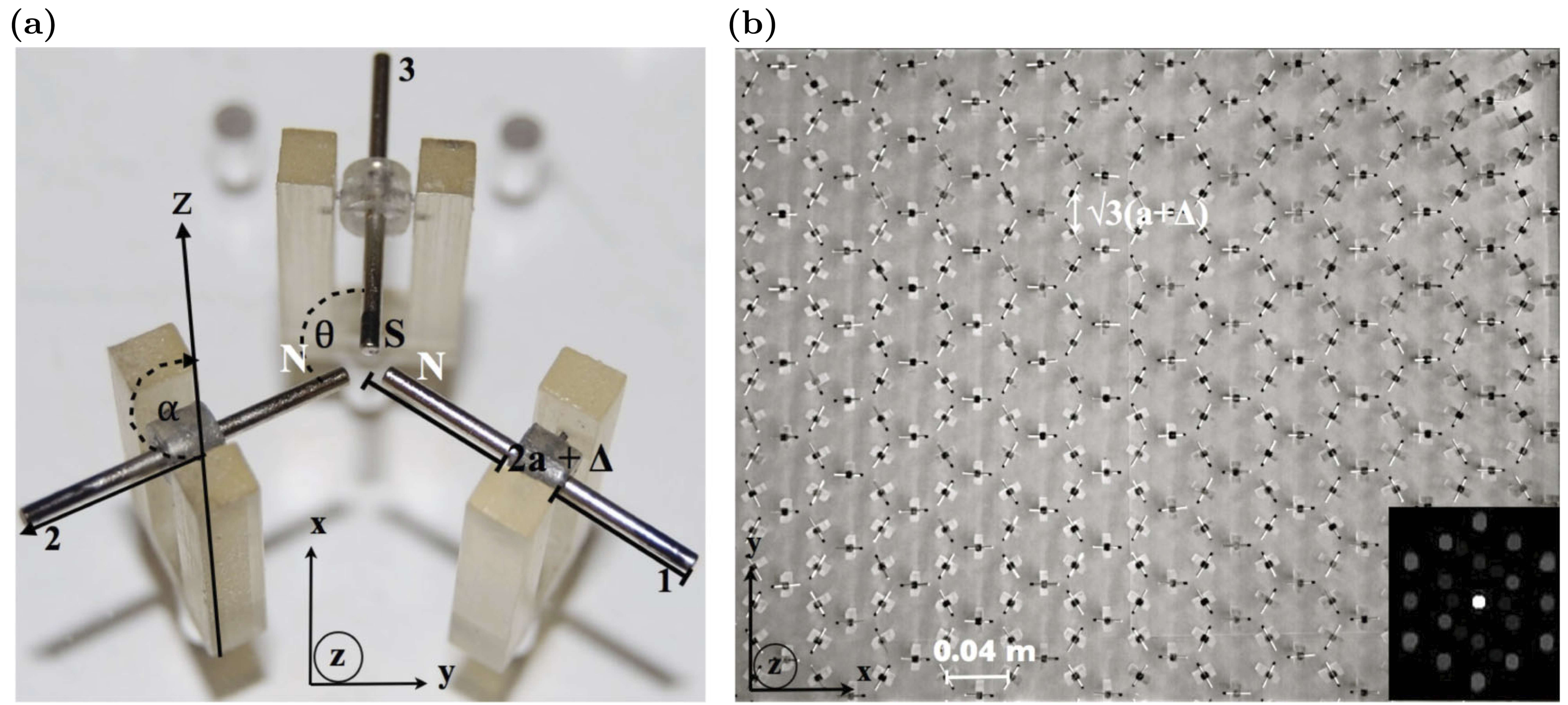}
\caption{(a) Three magnetic rotors 
composed by macroscopic ferromagnetic rods arranged 
at $\theta=120^{\circ}$ with respect to each other. (b)
Macosopic honeycomb lattice of rotors (inset shows the corresponding Fourier transform). Reprinted with permission
from \textcite{Mellado2012}.}
\label{fig:17}
\end{figure}

On a macroscopic scale, geometric frustration 
has been addressed by arranging classical bar magnets.
One significant case is shown in Fig.~\ref{fig:17}
where an ensemble of interacting millimeter-size magnets is arranged into a kagom{\' e} lattice.
The unit base of the system are 
ferromagnetic rods attached 
to planar rational units which allows  
only out of plane angular motion (polar angle $\alpha$) 
but not in plane one (the angle $\theta$ is fixed). The initial state of the system was 
prepared by polarizing these "macroscopic spins" 
along the perpendicular plane ($\hat{\bm{z}}$) via a strong static field. Switching off the field induces a relatively fast ($\sim 2$s) reorganization process, and the magnets stabilized into a 
equilibrium pattern filled by vertices with $2$-in/$1$-out 
and $1$-in/$2$-out. The rotors relaxation process occurred in three steps; namely a relatively fast inertial reorientation where the rotors break the axial symmetry, 
than a sequence of dissipative
librations followed by a final damped oscillations which leads to 
a nearest neighbor spin correlation $\langle \bm{s}_i \cdot \bm{s}_j \rangle = 1/3$. 
The authors also showed the exciting possibility to extend the 
system towards 3D, by stacking different plates composed by the rotors in a tetrahedral-like configuration. 
Such demonstration manifests the ubiquitous character of 
geometric frustration, which transcends length scale.

Numerous other condensed matter systems can be described as an assembly of particles with repulsive pairwise
interactions. Examples include
vortices in type-II superconductors \cite{libal2009},
ions \cite{Pyka2013}, 
dusty plasmas \cite{Morfill2009},
skyrmions \cite{ma2016}, and Wigner crystals \cite{Reichhardt2001a}.
Any of these systems, when coupled to the correct substrate geometry,
could exhibit effective spin degrees of freedom and ice rule.

\begin{figure}
  \begin{minipage}{\columnwidth}
  \includegraphics[width=\columnwidth]{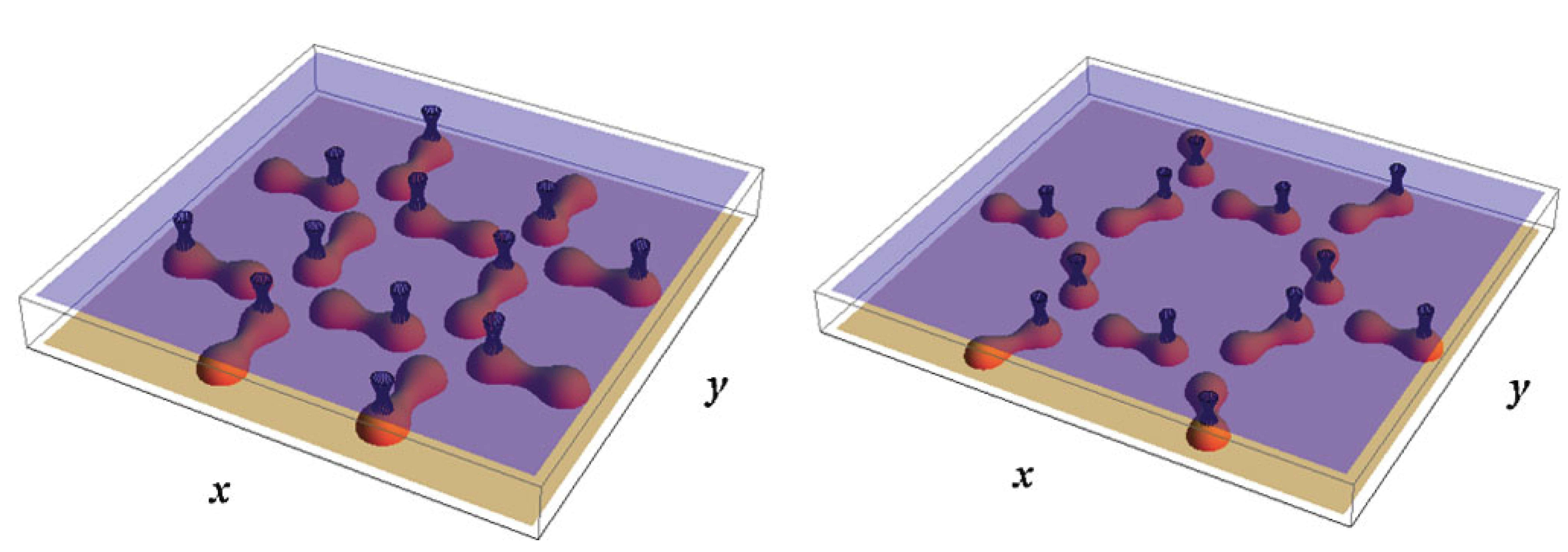}
  \end{minipage}
  \begin{minipage}{\columnwidth}
    \includegraphics[width=\columnwidth]{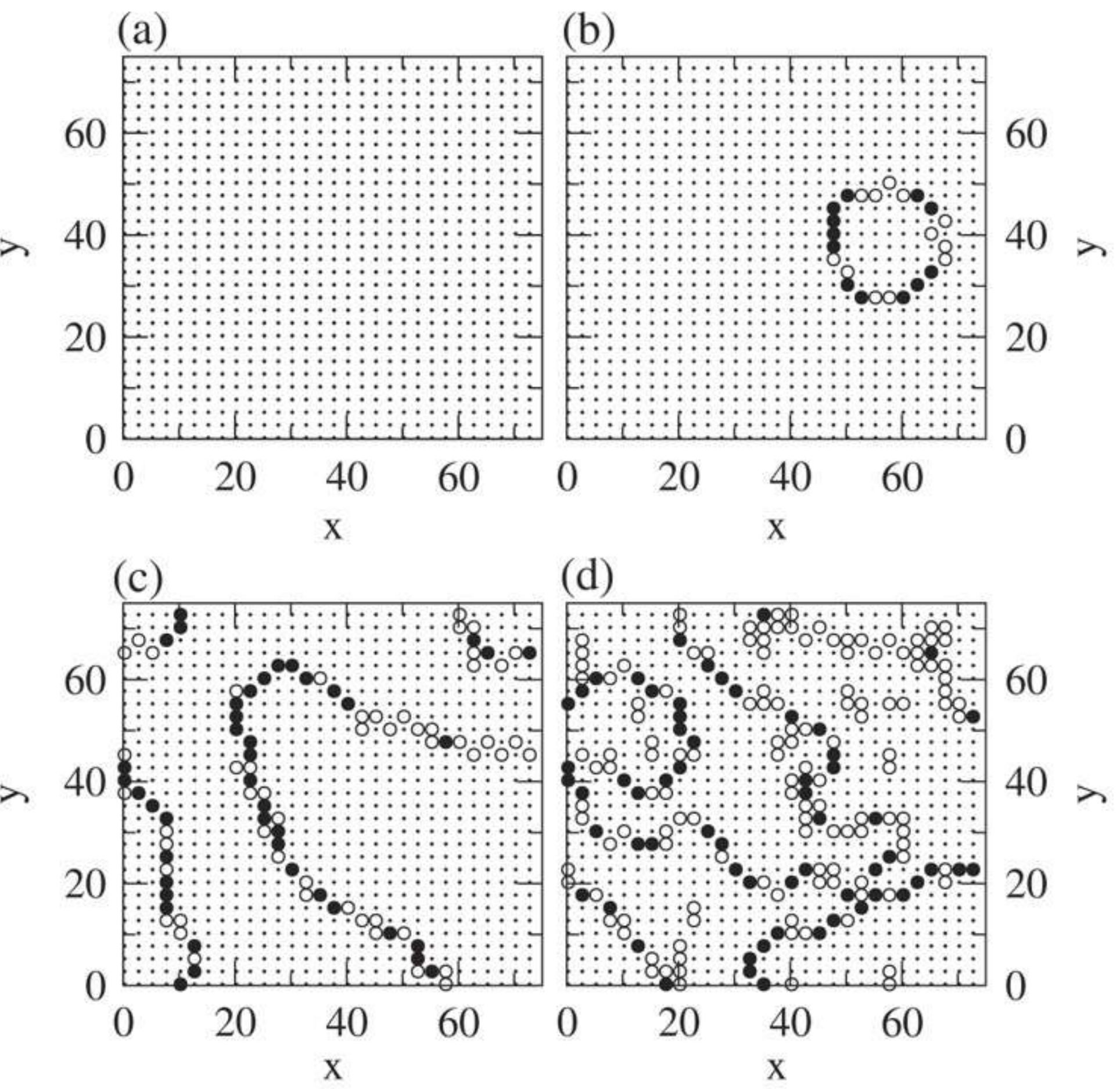}
    \end{minipage}
\caption{
Top panels:
  Schematic of the nanostructured pinning site configurations producing ice states.
  Double-lobed objects: pins; open mesh objects: vortices.
  Left: Square ice ground state. Right: One
  possible biased ground state of the kagom{\' e} ice system
\cite{libal2009}.
Grain boundary images in square ice samples 
for increasing disorder strength 
$\delta$.
Dots: ground state $n_{\rm in}=2$
ice-rule obeying
vertices; 
filled black circles: ice-rule defects; 
white circles: non-ice-rule defects.
(a) $\delta=0$. (b) $\delta=0.1$. (c) $\delta=0.5$.
(d) $\delta=1.0$.
Images from \textcite{libal2009}.
}
\label{fig:8}
\end{figure}

One of the first proposals for such particle-based artificial ices involved
vortices in type-II superconductors \cite{libal2009},
where considerable progress has already been made
in creating various types of pinning arrays to control the vortex ordering.  
In the vortex system,
a series of double well traps can be fabricated by placing
two pinning sites very close together,
as illustrated in the top panels of Fig.~\ref{fig:8}.
When the sample is nanostructured in this fashion, the thinner
parts of the superconductor have lower vortex condensation energy,
and the vortex will preferentially sit at the highest points of the double well
traps. 
For $N_p$ double well traps,
the number of vortices $N_{v}$ is directly proportional to the magnetic field,
so that when $N_{v}/N_{p} = 1/2$,
the system is equivalent to the colloidal square
ice at half filling.
Particle based modeling of  vortices in pinning arrays is performed
using similar techniques as those described for the colloidal systems;
however, the pairwise interaction between the vortices is a modified
Bessel function for vortices in a bulk superconductor, and has a $\ln(r)$ form
in thin film superconductors.
Many superconducting samples contain a significant amount of
intrinsic random disorder,
so \textcite{libal2009} explore the effects of additional quenched disorder
on the GS in a square ice pinning arrangement  
at $N_{v}/N_{p} = 1/2$, Fig.~\ref{fig:8}.
As the disorder strength $\delta$ increases, an increasing number of
non-ice rule obeying vertices appear in the GS in the form of
grain boundaries,
but individual monopoles that are not associated with a grain boundary
do not begin to appear until
$\delta \geq 0.5$.
The appearance of defects arranged in grain boundaries
was
subsequently observed experimentally in 
a square ASI \cite{Morgan2011}.

\begin{figure}
\includegraphics[width=\columnwidth]{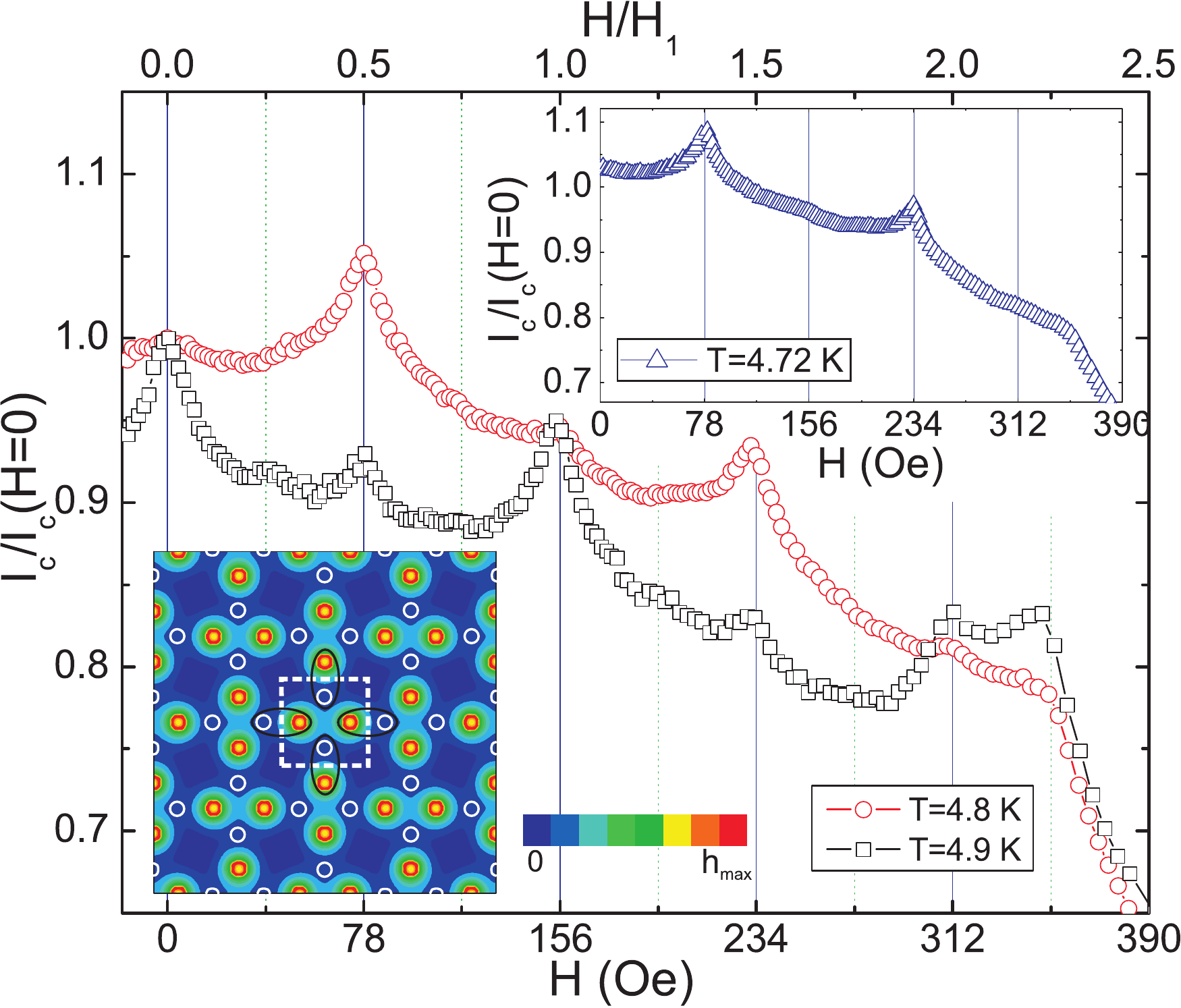}
\caption{
  The critical current of the vortex artificial square ice sample,
  normalized by the zero magnetic field critical current $I_c(0)$, as a
  function of the magnetic field
  $H$ at $T=4.8$ K (red circles)
  and $T=4.9$ K (black squares).
  Top inset shows the results obtained at
  $T=4.72$ K. The zero magnetic field critical currents are:
  $I_c(0)=802~\mu$A at $T=4.72$ K,
  $I_c(0)=613~\mu$A at $T=4.8$ K,
  and $I_c(0)=335~\mu$A at $T=4.9$ K.
  The lower inset shows contour plots
  of the simulated local magnetic field
  $h$ for the half matching field,
  which shows the ground state vortex configuration at
  $T=4.8$K.
  White circles indicate the positions of vortex-free holes
  and the dashed red square shows the unit cell of the simulation area.
Reprinted with permission from \textcite{Latimer2013}.
}
\label{fig:9}
\end{figure}

The superconducting vortex configurations can be observed directly
using imaging techniques, but it is also possible to probe
the stability of the vortex configurations
by applying a current in order to depin the vortices.
When the vortex arrangement is
highly ordered, the depinning threshold or critical
current $I_{c}$ is higher than
when the vortices are disordered.
Latimer {\it et al.} \cite{Latimer2013}
studied the critical currents of superconductors with square ice geometries
in experiments.
They find a series of peaks in the critical current $I_c$
as a function of vortex density.
When the temperature is lowered, a prominent critical current peak appears at 
$N_{v}/N_{p} = 1/2$,
as shown in Fig.~\ref{fig:9} for $I_{c}$ versus $H/H_{1}$, where
$H_{1}$ is the first matching field.
The inset shows a simulation of the flux
configuration at the $f=1/2$ filling where the square ice rule obeying state is highlighted. 
For $H/H_{1} > 1.0$, the vortices start to become doubly quantized, 
and another peak in the critical current appears at $H/H_{1} = 3/2$,
which again corresponds 
to an ice rule obeying GS.  
 
In a series of experiments, \textcite{Trastoy2014a,Trastoy2015} examined spin ice
pinning arrays
for the high temperature superconductor
YBCO, where thermal effects are important.
They found that changes in resistance correspond to
vortex configurations that are
dominated by a geometrically frustrated energy landscape
that favors ice like ordering and frustration,
and they find a series of peaks in the transport response
similar to what is found in ordered square pinning lattice arrangements.

For vortices in a kagom{\' e} pinning arrangement,
various ordered and disordered arrangements have been observed
in 
imaging experiments, and it has been argued
that the long-range
vortex-vortex interactions
are insufficient to lift the degeneracy of the
different vortex configurations in the strong pinning sites \cite{Xue2017}.
As the magnetic field is increased,
additional vortices become trapped in the interstitial regions surrounding
the pinned vortices,
which could be a step toward the realization of a stuffed artificial spin ice \cite{Xue2018}.   
Imaging experiments for vortex configurations on the
kagom{\' e} ice \cite{Wang2018}
showed the predicted 
kagom{\' e} ice rule obeying states \cite{libal2009} both at $H/H_{1} = 1/2$, as
expected,
and also 
at the higher field of $H/H_{1} = 3/2$,
where the additional vortices occupy the interstitial regions between pinning sites.
Interestingly,
the ice rule state
at $H/H_{1} = 3/2$
is even
more ordered than the state at $H/H_{1} = 1/2$,
suggesting that the interstitial vortices may play a role in annealing defects in
the ice GS.

\begin{figure}
\includegraphics[width=\columnwidth]{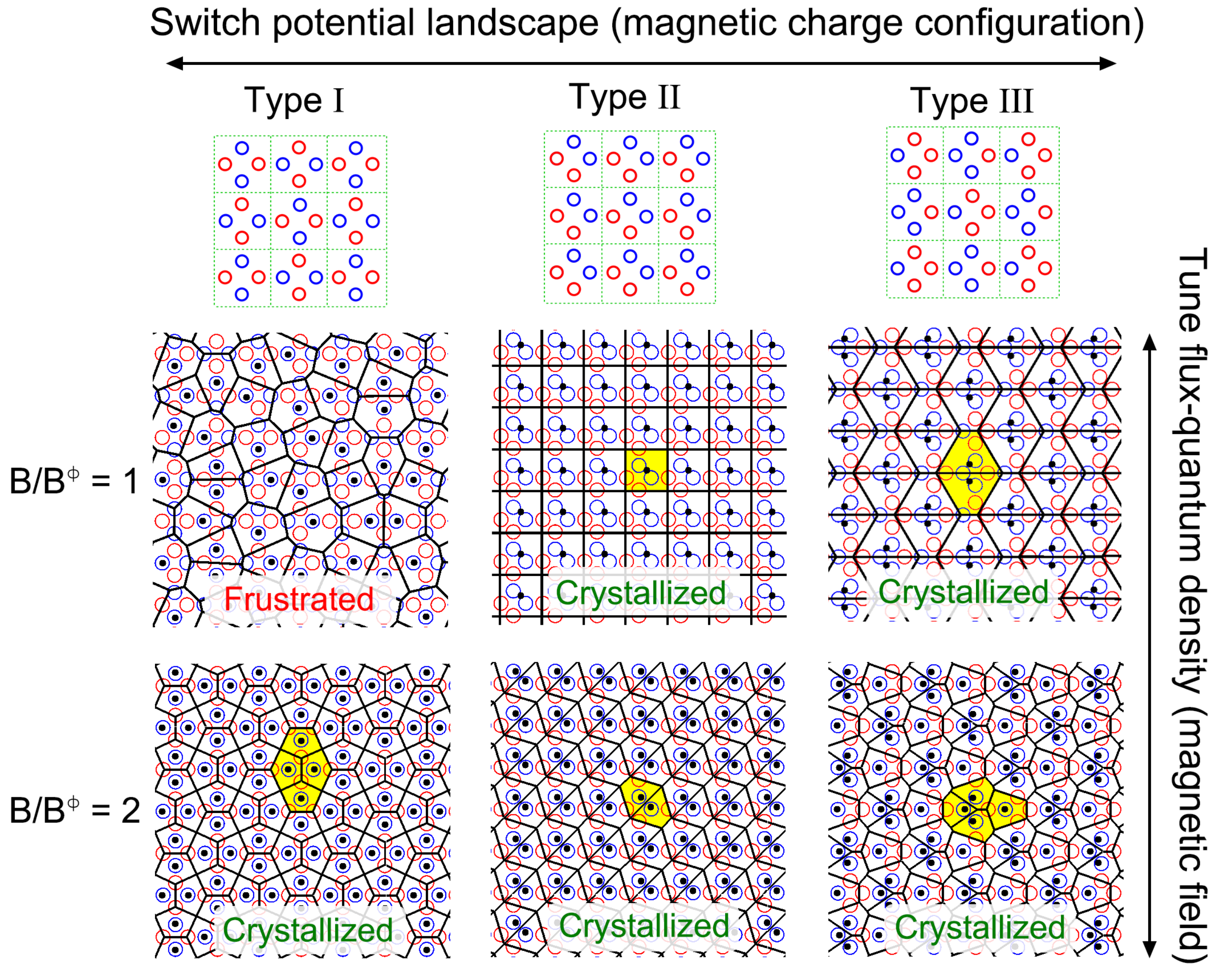}
\caption{
  Simulated distributions of flux quanta under type I, II and III magnetic-charge orders and various flux-quantum densities. Solid black lines indicate Voronoi diagrams clarifying the flux-quantum ordering. One repeating unit structure in each of the crystallized flux-quantum lattices is highlighted in yellow. The disordered flux quanta under the type I magnetic-charge configuration at $B/B_\phi$ = 1 originate from geometric frustration with very high degeneracy.
Reprinted with permission
from the authors \textcite{Wang2018}.}
\label{fig:10}
\end{figure}

Wang {\it et al.} \cite{Wang2016b} examined vortex ordering and dynamics on
a switchable artificial spin ice array that was motivated by the 
realization of a magnetic rewritable artificial spin ice \cite{Ge2018}.
In this system,
frustration effects can be switched on or off by changing the magnetic charge ordering 
of the underlying reconfigurable magnetic ice substrate.
Different magnetic charge orderings can 
produce disordered or frustrated vortex configurations,
as well as non-frustrated configurations
that produce ordered vortex crystals,
as illustrated in Fig.~\ref{fig:10} for different possible ordered states
at $B/B_{\phi} = 1.0$ and the $f=2.0$ filling.
Here types I, II and III correspond to different magnetic charge ordered
states of the magnetic substrate \cite{Wang2018}. Other work on vortex states with 
rewritable magnetic ice substrates
showed that the different patterns can be switched with an applied current \cite{Ge2018}. 

Square and kagom{\' e} spin ice geometries have also been proposed for
skyrmions in magnetic dot arrays,
where the dots are arranged such
that the skyrmion can sit on either side of the dot.
A recent work demonstrated that, 
when the skyrmions are strongly localized or particle like, the behavior of the system
is very similar to what is observed in both
superconducting vortex and colloidal artificial ices \cite{ma2016}.
When the magnetic field is varied, however,
the skyrmions can become elongated and fill the entire dot, destroying the
frustration,
so the system can exhibit a transition from 
an ice rule obeying state to square or triangular ordered states \cite{ma2016}.

\section{Outlook}
In this section we provide 
further general directions 
that cover different disciplines across condensed matter systems 
not limited to particle based systems.

\textbf{Stuffed Spin Ice}: 
In particle based artificial ices, additional particles can be placed in the
spaces between the traps of the substrate as well as within the traps themselves.
These interstitial particles could potentially modify the ice rule for the
particles that are sitting in the trap sites.
One study has already provided evidence
that the interstitial particles
can actually enhance the ordering of the ice state \cite{Xue2018}.
It would also be interesting to explore
the dynamics of
interstitial particles for
different ice substrate configurations.
For example, an ordered square ice background might produce
increased or decreased diffusion of the interstitial particles compared
to a degenerate GS ice background.\\
\\
\textbf{Driven Dynamics}:
In ASI
the spin degrees of freedom are permanently localized on the
magnetic nanoisland.
In contrast,
in particle based artificial ices
it is possible for the particles to hop from one ice substrate trap to another,
making it possible to explore
the dynamics of a frustrated system
compared to that of a
non-frustrated one.
Protocols that could be considered include
sliding dynamics under a dc drive,
shaken dynamics under external forces,
or simply applying a drive to only one side of the sample but not to the other.
This would be equivalent to subjecting
a portion of an
ASI to a magnetic field while the
other portion of the sample experiences no field.\\
\\
\textbf{Dynamic Substrates}:
If optical traps are used to create the ice substrate,
than a new type of dynamics involving the traps themselves could be implemented.
For example,
the trap could be flashed on and off, or the orientation of the double well potential
could be rotated.
Such protocols could
provide new methods
to reduce frustration effects.
Alternatively,
it may be possible to induce new types of frustration effects
using 
dynamical protocols 
that bring repulsive particles together.
Such protocols would
create local high energy states,
but due to the frustration, there may 
not be an easy direction in which one of the particles could move away
in order to reduce the interaction energy. 
In particle  based systems, dynamic protocols could
be introduced readily by using dynamic traps \cite{Curtis2002,Bhebhe2018}.\\ 
\\
\textbf{Frustrated Active Matter Systems}:
Active matter composed of self-propelled particles
is a rapidly growing research field, with an explosion
in the experimental realization of artificially self-propelled colloids \cite{Bechinger2016b}.
An interesting area for future exploration is
frustrated systems on lattices where the fluctuations are active rather than thermal.
Some studies have already shown the organization of active matter
into vortex states \cite{Beppu2017} and vortex lattices \cite{Wioland2016,Nishiguchi2018}, and it would
be possible to arrange these vortices into a frustrated configuration.
Additionally, if active and passive particles are mixed together,
a frustrated ice geometry
might
reduce the diffusion of the active particles,
while an ordered ice state could produce flow paths for the active particles.
Other systems such as coupled gyroscopes or spinners
could be placed in frustrated geometries 
which could lead to interesting dynamics \cite{Nash2015}.\\
\\
\textbf{Deformable Substrates}:
In most of the ASI studied
to date, the substrate or confinement is fixed and cannot react to the 
forces exerted on it by the particles.
In soft matter systems there are many ways to create
a substrate that could react and be deformed elastically.
If this reaction is introduced into an ASI system,
it could lead to new methods for reducing frustration effects.
In addition, excitations such as monopoles could create
additional long range strain fields in the soft substrate that
could be attractive or repulsive for other monopoles,
leading to emergent effects
such as excitons and polarons.\\     
\\
\textbf{Nonlocal Frustration Effects}:
Numerous experiments have shown that large numbers of
colloids can be addressed individually in optical
feedback experiments \cite{Lavergne2019}.
Thus, another route for studying
frustrated systems would be to
introduce non-local effects in which the forces experienced by particles in
one part of the system are correlated with the positions of other particles that are
far away.
This would make it possible to test whether
phase
transitions can be induced by increasing the strength of the nonlocal interactions,
to study the effects of small world versus random interactions, and to determine
whether nonlocal effects change the nonequilibrium properties of the frustrated
state.
It would also be possible to engineer
nonlocal frustration
effects in which a reduction of frustration at one location in the system increases
the frustration at a distant point, and to study how the system would
relief such effects.\\
\\
\textbf{Continuous Spin Directions}:
Another avenue in 2D soft mater systems
is to allow the effective spin degree of freedom to point along multiple directions
or have some freedom to move in the third dimension
in order to create
a frustrated Potts model, 
Heisenberg like spin models, and/or XY models.\\
\\
\textbf{Nondissipative Dynamics}:
In the particle based models considered so far,
the motion of the particles is overdamped.
It is, however, possible to realize
particle ices in which nondissipative effects are important,
such as in underdamped systems including
dusty plasmas, trapped ion crystals, or even milimetric magnets on 3D printed substrates.
The inertial effects could produce phononic excitations, and it would be possible to study the difference in the
phonon modes for
crystalline versus degenerate or frustrated ice states, as well as
the propagation of solitons or shock waves.\\
\\
\textbf{Functionalized Colloids for Frustrated Geometries}:
There are many examples of functionalized colloidal systems in which
the interactions between the colloids can be tailored
to be anisotropic \cite{Glotzer2007}, reconfigurable \cite{Ortiz2014a}, or patchy \cite{Bianchi2011},
or even to mimic chemical bonds \cite{Wang2012}.
Using such systems,
it would be possible to create 2D or 3D colloidal assemblies
that naturally exhibit frustration, or even 
a colloidal system that mimics the structure of water ice \cite{Pauling1935}.\\
\\
\textbf{Quantum Effects}:
A final potential avenue would be to place cold atoms or cold ions on trap arrays
arranged in an ice structure.  This system could be realized at a size scale and
temperatures for which
quantum effects could become important
\cite{Glaetzle2015,Mazurenko2017}.

\begin{acknowledgments}
We acknowledge helpful discussions with S. T. Bromley, B. Canals, J. Casademunt, L. Cugliandolo, T. M. Fischer, G. Franzese, S. Giblin, S. Klapp, D. Levis, A. Libal, H. L\"owen, E. Oguz, J. Ort\'in, I. Pagonabarraga, N. Rougemaille, F. Sagu\'es, J. M. Sancho, A. Snezhko, Y. Shokef, A. Straube, E. Vives, Z. Xiao and W. Yong-Lei.
A. O. A. and P. T. acknowledges support from
the ERC Consolidator grant ``ENFORCE'' (No. 811234).
The work of C. R., C. J. O. R., and C. N. was carried out under the auspices of the US Department of Energy through
the Los Alamos National Laboratory.  Los Alamos National Laboratory is
operated by Triad National Security, LLC, for the National Nuclear Security
Administration of the U. S. Department of Energy (Contract No. 892333218NCA000001).
P. T. acknowledges support from from
MINECO 344 (FIS2016-78507-C2), DURSI (2017SGR1061)
and Generalitat de Catalunya under Program ``ICREA
Acad\'emia".
\end{acknowledgments}
\clearpage
\bibliography{biblio}
\end{document}